\documentclass[nofootinbib,prd,aps,superscriptaddress,preprintnumbers,twocolumn,showpacs]{revtex4-1}
\usepackage[dvipdfmx]{graphicx}
\usepackage{amsmath,amssymb,amsthm,bm,bigdelim,multirow}
\usepackage{color}

\newcommand{\kslash}{k\kern-1ex /}
\newcommand{\pslash}{p\kern-1ex /}
\newcommand{\qslash}{q\kern-1ex /}
\newcommand{\lslash}{l\kern-1ex /}
\newcommand{\sslash}{s\kern-1ex /}
\newcommand{\Dslash}{D\kern-1.2ex /}

\newcommand{\beqa}{\begin{eqnarray}}
\newcommand{\eeqa}{\end{eqnarray}}

\newcommand{\bd}{\begin{description}}
\newcommand{\ed}{\end{description}}

\newcommand{\ben}{\begin{eqnarray}}
\newcommand{\een}{\end{eqnarray}}
\newcommand{\nn}{\nonumber}
\def\lsim{\raise0.3ex\hbox{$<$\kern-0.75em\raise-1.1ex\hbox{$\sim$}}}
\def\gsim{\raise0.3ex\hbox{$>$\kern-0.75em\raise-1.1ex\hbox{$\sim$}}}
\def\simgt{\rlap{\lower 3.5 pt\hbox{$\mathchar \sim$}}\raise 2.0pt \hbox {$>$}}
\def\simlt{\rlap{\lower 3.5 pt\hbox{$\mathchar \sim$}}\raise 2.0pt \hbox {$<$}}

\newcommand{\crd}{\color{black}}

\begin{document}
\title{Finite size effect on vector meson and baryon sectors in 2+1 flavor QCD\\ at the physical point}

\author{K.-I.~Ishikawa}
\affiliation{Graduate School of Science, Hiroshima University, Higashi-Hiroshima, Hiroshima 739-8526, Japan}

\author{N.~Ishizuka}
\affiliation{Center for Computational Sciences, University of Tsukuba, Tsukuba, Ibaraki 305-8577, Japan}

\author{Y.~Kuramashi}
\affiliation{Center for Computational Sciences, University of Tsukuba, Tsukuba, Ibaraki 305-8577, Japan}

\author{Y.~Nakamura}
\affiliation{RIKEN Center for Computational Science, Kobe, Hyogo 650-0047, Japan}

\author{Y.~Namekawa}
\affiliation{High Energy Accelerator Research Organization (KEK), Tsukuba, Ibaraki 305-0801, Japan}

\author{E.~Shintani}
\affiliation{RIKEN Center for Computational Science, Kobe, Hyogo 650-0047, Japan}

\author{Y.~Taniguchi}
\affiliation{Center for Computational Sciences, University of Tsukuba, Tsukuba, Ibaraki 305-8577, Japan}

\author{N.~Ukita}
\affiliation{Center for Computational Sciences, University of Tsukuba, Tsukuba, Ibaraki 305-8577, Japan}

\author{T.~Yamazaki}
\affiliation{Faculty of Pure and Applied Sciences, University of Tsukuba, Tsukuba, Ibaraki 305-8571, Japan}

\author{T.~Yoshi\'e}
\affiliation{Center for Computational Sciences, University of Tsukuba, Tsukuba, Ibaraki 305-8577, Japan}

\collaboration{PACS Collaboration}

\begin{abstract}
We investigate the finite size effect on the vector meson and the baryon sectors using a subset of the ``PACS10'' configurations which are generated, keeping the space-time volumes over (10 fm$)^4$ in 2+1 flavor QCD at the physical point. 
Comparing the results on (5.5 fm$)^4$ and (10.9 fm$)^4$ lattices, the ground states of octet baryons , which are stable on the lattice, show no finite size effect within less than 0.5\% level of statistical errors. For those of vector mesons, which are unstable on the lattice, we observe that the effective masses are well below the experimental resonance levels both on (5.5 fm$)^4$ and (10.9 fm$)^4$ lattices. 
For the decuplet baryon sector, we have found that the time dependence of the effective mass looks quite similar to that for the vector meson sector, including the $\Omega$ baryon channel. We discuss its origin due to a possible mixing with the nearby multihadron states. Since the $\Xi$ baryon mass can be determined with the smallest ambiguity among the vector meson and the baryon masses, we use it together with the pion and kaon masses as the physical inputs to determine the physical point.

\end{abstract}
\date{\today}

\preprint{UTHEP-733, UTCCS-P-123, HUPD-1909}

\maketitle

\newpage
\section{Introduction}

The ``PACS10'' configurations  generated by PACS Collaboration have a physical lattice size over (10 fm$)^4$ at the physical point in 2+1 flavor QCD~\cite{mf_nf2+1}. This is the successor to the PACS-CS project to reduce the up-down quark masses down to the physical point with the Wilson-type quark action~\cite{pacs-cs1,pacs-cs2}. There are inherent advantages of large scale simulations: the statistical error can be reduced with the optimized use of the geometrical symmetries~\cite{master-field,mf_nf0,ukita18}, and the allowed minimum momentum and the resolution diminish in proportion to $1/L$ with $L$ as the lattice extent as we have exploited for the calculation of the nucleon charge radius~\cite{nff_128} and the hadron vacuum polarization (HVP) contribution to the muon $g-2$~\cite{g-2_128}. 

The PACS10 configurations also give us a good opportunity to investigate the finite size effect at the physical point in 2+1 flavor QCD. In our previous work~\cite{mf_nf2+1}, a finite size study for the pseudoscalar (PS) meson sector employing (10.9 fm$)^4$ and (5.5 fm$)^4$ lattices has been carried out at a cutoff of $a^{-1}\approx 2.3\ $GeV. We have employed two kinds of analysis. One is a comparison between the results  at the same hopping parameters on both lattices. The other is a comparison at the same axial Ward identity (AWI) quark masses, $m_{\rm ud}^{\rm AWI}$, on both lattices, where the quark masses on the smaller lattice is adjusted to those on the larger one by the reweighting method~\cite{pacs-cs2}. The former analysis revealed 2.1(8)\%, 4.8(1.6)\%, and 0.36(31)\% finite size effect on $m_\pi$, $m_{\rm ud}^{\rm AWI}$, and $f_\pi$, respectively. On the other hand, in the latter analysis, we detect 0.66(33)\%, 0.26(13)\%, and 0.40(32)\% finite size effect on $f_\pi$, $f_K$, and $f_K/f_\pi$, respectively. These values are {\crd roughly} 3 times larger than the chiral perturbation theory (ChPT) predictions of 0.20\%, 0.08\%, and 0.13\% for $f_\pi$, $f_K$, and $f_K/f_\pi$~\cite{fse_chpt}. In our recent study of HVP in the muon $g-2$, we have also found a roughly twice larger finite size effect compared to the ChPT prediction~\cite{g-2_128}. 

In this paper, we investigate the finite size effect for the vector meson and the baryon sectors on the same gauge configurations in Ref.~\cite{mf_nf2+1}, employing the all-mode-averaging (AMA) technique \cite{Shintani:2014vja,vonHippel:2016wid} and the ``crystal'' source method, which is a sort of ${\mathbb Z_3}$ grid source method \cite{Li:2010pw,Gong:2013vja,Wu:2018tvt}, to obtain hadron propagators with high precision. With the use of  $m_\pi$, $m_K$, and $m_\Xi$ as the physical inputs, we determine the physical point together with the lattice cutoff.

This paper is organized as follows. In Sec.~\ref{sec:simulation}, we explain the simulation details, including the numerical technique, and discuss the possible mixing states in the vector meson and the baryon sectors on the lattice. Section~\ref{sec:results} presents the results for the finite size effect, and in Sec.~\ref{sec:physicalpoint}, we discuss the physical point with the use of $m_\pi$, $m_K$, and $m_\Xi$ as the physical inputs. Our conclusions and outlook are summarized in Sec.~\ref{sec:conclusion}.

\section{Simulation details}
\label{sec:simulation}
\subsection{PACS10 gauge configurations}
\label{subsec:lattice_param}
The finite size effect for the vector meson and baryon sectors are studied by employing the same subset of PACS10 gauge configurations as for the pseudoscalar meson sector in Ref.~\cite{mf_nf2+1}. Since the details of the configuration generation have been already given in Ref.~\cite{mf_nf2+1}, we briefly present only the relevant points to make this paper self-contained.  

The configurations were generated with the stout-smeared $O(a)$-improved Wilson-clover quark action and Iwasaki gauge action~\cite{iwasaki} on $V=L^3\times T=128^4$ and $64^4$ lattices at $\beta=1.82$ in 2+1 flavor QCD. We use the stout smearing parameter $\rho = 0.1$~\cite{Morningstar:2003gk}, and the number of the smearing steps is six. The nonperturbative improvement coefficient of $c_{\rm SW}=1.11$ is determined by the Schr{\"o}dinger functional (SF) scheme~\cite{csw_sf}. The hopping parameters are chosen to be $(\kappa_{\rm ud},\kappa_{\rm s})=(0.126117,0.124902)$ for the light (up-down) and strange quarks. Hereafter, this combination of hopping parameters is called the original hopping parameters at the simulation point. The physical lattice sizes are (10.9 fm$)^4$ and (5.5 fm$)^4$, respectively, using the lattice spacing of $a=0.08520(16)$ fm [$a^{-1}=2.3162(44)$ GeV] determined by the precision measurement of the baryon spectroscopy, which will be explained in Sec.~\ref{sec:physicalpoint}. After thermalization, we generate 2000 and 200 trajectories on $64^4$ and $128^4$ lattices, respectively, and calculate hadronic observables at every ten trajectories. 
The statistical errors are estimated with the jackknife method.
After investigating the bin size dependence, we have
chosen 50 and 10 trajectories for $64^4$ and $128^4$ lattices,
respectively.

When we investigate the finite size effect equalizing the AWI quark masses between $64^4$ and $128^4$ lattices, we need to make a tiny shift of the hopping parameters on the smaller lattice employing the reweighting technique~\cite{mf_nf2+1}. This technique also allows us to determine the quark mass dependence of the hadron masses, which is used in Sec.~\ref{sec:physicalpoint}. We choose five combinations of the reweighted hopping parameters: $(\kappa_{\rm ud}^*,\kappa_{\rm s}^*)$=$(0.126117,0.124882)$, $(0.126117,0.124922)$, $(0.126119,0.124882)$, $(0.126119,0.124902)$, $(0.126119,0.124922)$.  The reweighting factors have been already evaluated in Ref.~\cite{mf_nf2+1} following the techniques presented in Refs.~\cite{pacs-cs2, hasenfratz, RBC_UKQCD}. 

\subsection{Precise measurement of hadron correlators}
\label{subsec:measurement}

\subsubsection{Definition of hadron operators}
We measure the vector meson and baryon correlators employing appropriate operators. The vector meson operators are expressed as
\ben
M_\mu^{fg}(x)={\bar q}_f(x)\gamma_\mu q_g(x),
\een
where $f$ and $g$ denote quark flavors and $\gamma_\mu$ $(\mu=1,2,3,4)$ are Dirac matrices.

The octet baryon operators are given by 
\ben
{\cal O}^{fgh}_\alpha(x)=\epsilon^{abc}((q_f^a(x))^T C\gamma_5 q_g^b(x))
q_{h\alpha}^c(x),
\een 
where $a,b,c$ are color indices, $C=\gamma_4\gamma_2$ is the charge conjugation matrix, and $\alpha=1,2$ labels the $z$ component of the spin 1/2 (in Dirac representation $\alpha$ consists of four spinor components). Referring to Ref.~\cite{Aoki:2002fd}, the $\Sigma$- and $\Lambda$-like octet baryons are distinguished by the flavor structures,
\ben
\Sigma{\rm -like}\;\; &:& \;\; -\frac{{\cal O}^{[fh]g}+{\cal O}^{[gh]f}}{\sqrt{2}},\\
\Lambda{\rm -like}\;\; &:& \;\; \frac{{\cal O}^{[fh]g}-{\cal O}^{[gh]f}-2{\cal O}^{[fg]h}}
{\sqrt{6}},
\een
where ${\cal O}^{[fg]h}={\cal O}^{fgh}-{\cal O}^{gfh}$. The decuplet baryon operators for the four $z$ components of the spin 3/2 is defined as
\ben
D^{fgh}_{3/2}(x)&=&-\epsilon^{abc}((q_f^a(x))^T C\Gamma_+
q_g^b(x))q_{h1}^c(x),\label{eq:dec1}\\
D^{fgh}_{1/2}(x)&=&
\epsilon^{abc}[((q_f^a(x))^T C\Gamma_0q_g^b(x))q_{h1}^c(x)\nn\\
&&-((q_f^a(x))^T C\Gamma_+q_g^b(x))q_{h2}^c(x)]/\sqrt{3},\label{eq:dec2}\\
D^{fgh}_{-1/2}(x)&=&
\epsilon^{abc}[((q_f^a(x))^T C\Gamma_0q_g^b(x))q_{h2}^c(x)\nn\\
&&+((q_f^a(x))^T C\Gamma_-q_g^b(x))q_{h1}^c(x)]/\sqrt{3},\label{eq:dec3}\\
D^{fgh}_{-3/2}(x)&=&\epsilon^{abc}((q_f^a(x))^T C\Gamma_-
q_g^b(x))q_{h2}^c(x),\label{eq:dec4}
\een
where $\Gamma_{\pm}=(\gamma_1\mp i\gamma_2)/2$, $\Gamma_0=\gamma_3$ and the flavor structures should be symmetrized. 
The decuplet baryon correlator is obtained by the summation over four $z$ components as, 
\begin{eqnarray}
 && \sum_{\vec x}
 \sum_{S=\pm3/2,\pm1/2}
 \left[
 \langle D_{S}^{fgh}(x) \overline{D}_{S}^{fgh}(0)\rangle
 \right]
\end{eqnarray}
with operators in Eqs.~(\ref{eq:dec1})-(\ref{eq:dec4}). $\sum_{\vec x}$ represents the summation over the spatial lattice site ${\vec x}=(x_1,x_2,x_3)$.

{\crd It can be easily found that the correlators based on the decuplet baryon operators $D_{\pm\frac{3}{2},\pm\frac{1}{2}}^{fgh}(x)$ are equivalent to those constructed using the vector-spinor interpolating operators $B_{\mu\beta}(x)\ (\beta=1,2,3,4)$ projected to the spin-parity of $j^P=\frac{3}{2}^+$ \cite{Ioffe:1981kw}. For example, 
the equivalence of the $\Omega$ baryon correlator is written as
\begin{eqnarray}
 && \sum_{\vec x}
 \sum_{S=\pm3/2,\pm1/2}
 \left[
 \langle D_{S}^{sss}(x) \overline{D}_{S}^{sss}(0)\rangle
 \right]\nonumber\\
&& =
\sum_{\vec x}
\sum_{\mu,\nu=1,2,3,4}
{\rm tr} \left[P^{\frac{3}{2}^+}_{\mu\nu}
\langle
 B_{\nu}^{\Omega}(x) \overline B_{\mu}^{\Omega}(0)
\rangle\right]
\label{equiv_corr}
\end{eqnarray}
with 
\begin{eqnarray}
 B_{\mu\beta}^{\Omega}(x) &=& \epsilon^{abc}((s^a(x))^T C\gamma_\mu
s^b(x))s_\beta^c(x) / \sqrt{2},
\end{eqnarray}
where 
${\rm tr}$ denotes the trace over spinor indices and the spin-parity projection operator $P_{\mu\nu}^{\frac{3}{2}^+}$ is defined by $P_{\mu\nu}^{\frac{3}{2}^+}=\left(\frac{1+\gamma_4}{2}\right)\left(\delta_{\mu\nu}-\frac{1}{3}\gamma_\mu\gamma_{\nu}\right)$ for $\mu,\nu=1,2,3$ and $P_{44}^{\frac{3}{2}^+}=0$ for the zero spatial momentum state.
The vector-spinor operators for other decuplet baryons $\Delta, \Sigma^*$, $\Xi^*$ are defined as
\begin{eqnarray}
 B_{\mu\beta}^{\Delta}(x) &=&
 \epsilon^{abc}((u^{a}(x))^TC\gamma_{\mu}u^b(x))u_{\beta}^c(x) / \sqrt{2},
 \\
 B_{\mu\beta}^{\Sigma^*}(x) &=&
 \epsilon^{abc}
 [
  2((u^{a}(x))^TC\gamma_{\mu}s^b(x))u_{\beta}^c(x)\nonumber \\&&
  +((u^{a}(x))^TC\gamma_{\mu}u^b(x))s_{\beta}^c(x)
 ] / \sqrt{6},
 \\
 B_{\mu\beta}^{\Xi^*}(x) &=&
 \epsilon^{abc}
 [
  2((u^{a}(x))^TC\gamma_{\mu}s^b(x))s_{\beta}^c(x)\nonumber \\&&
  +((s^{a}(x))^TC\gamma_{\mu}s^b(x))u_{\beta}^c(x)
 ] / \sqrt{6}.
\end{eqnarray}
The equivalence as shown with Eq.~(\ref{equiv_corr}) also holds for $\Delta, \Sigma^*$, and $\Xi^*$. 
}

\subsubsection{All-mode-averaging}
{\crd In order to reduce the statistical errors of hadron propagators, especially for the baryon sector, we employ the all-mode-averaging (AMA) technique~\cite{Shintani:2014vja}. On this technique, the parameters for approximation, which is obtained from the low-precision solver with the truncated iteration $N_{\rm iter}^{f}$ at quark flavor $f$, is tuned to be strong correlation with the high-precision solver, but its computational cost is much lower than high-precision solver. Here, the approximation is constructed with the same procedure as in Ref.~\cite{vonHippel:2016wid}, in which the Schwartz alternative procedure (SAP) through domain decomposition is used for the preconditioning and approximation. In addition, the deflation field method with $N_{\rm defl}$ fields is used to construct the approximation efficiently for the light quark via the low-mode projection. For the strange quark, we use the SAP without a deflation field to save the memory size and avoid the computation time of a low-mode projection. The solver algorithm we employed here is a general conjugate residual (GCR). Our truncated iteration and the domain size on 64$^4$ and 128$^4$ lattices are listed in Table~\ref{tab:AMA_parameter}. 
  \begin{table}
    \begin{center}
      \caption{AMA parameters on $128^4$ and $64^4$ lattices. The limited iteration of the solver denotes $N_{\rm iter}^{\rm ud}$ for a light quark and $N_{\rm iter}^{\rm s}$ for a strange quark. $N_{\rm defl}$ denotes the number of deflation fields.}\label{tab:AMA_parameter}
      \begin{tabular}{ccc}
        \hline\hline
        & {$128^4$ lattice} & {$64^4$ lattice}\\
        \hline
($N_{\rm iter}^{\rm ud}$,$N_{\rm defl}$) [GCR$_{\rm SAP+defl}$] & (12,38) & (5,28) \\
$N_{\rm iter}^{\rm s}$ [GCR$_{\rm SAP}$] & 9 & 10 \\
Domain size &  $4^4$ & $8^4$ \\
Total measurements & 7679 & 25573\\  
High precision measurements & 4/config. & 1/config.\\    \hline\hline        
      \end{tabular}
    \end{center}
  \end{table}
}
  
The vector meson and the baryon correlators are calculated with the smeared source and the local sink. The smeared source is constructed with an exponential smearing function $\Psi(|{\vec x}|)=A_q\exp(-B_q|{\vec x}|)$ $(q={\rm ud,s})$, where $\Psi(0)=1$ for the ud and s quark propagators. We adjust the parameters such that the effective mass for the octet baryon should reach a plateau quickly. Our choice is $A_{\rm ud}=1.2, B_{\rm ud}=0.14$ and $A_{\rm s}=1.2, B_{\rm s}=0.21$ for both the $64^4$ and $128^4$ lattices.

Note that the details of the calculation of AWI quark masses and the PS meson masses are described in Ref.~\cite{mf_nf2+1}, where the wall source method without gauge fixing~\cite{ws_ngf} is employed to measure the correlators with the use of the local PS operator and the local axial vector current. The solver algorithm is the mixed precision BiCGStab method with the tolerance of $|Dx-b|/|b|<10^{-8}$.

\subsubsection{Crystal source method}
\label{sec:crystal_source}

\begin{figure}[t]
\begin{center}
\includegraphics[width=90mm,angle=0]{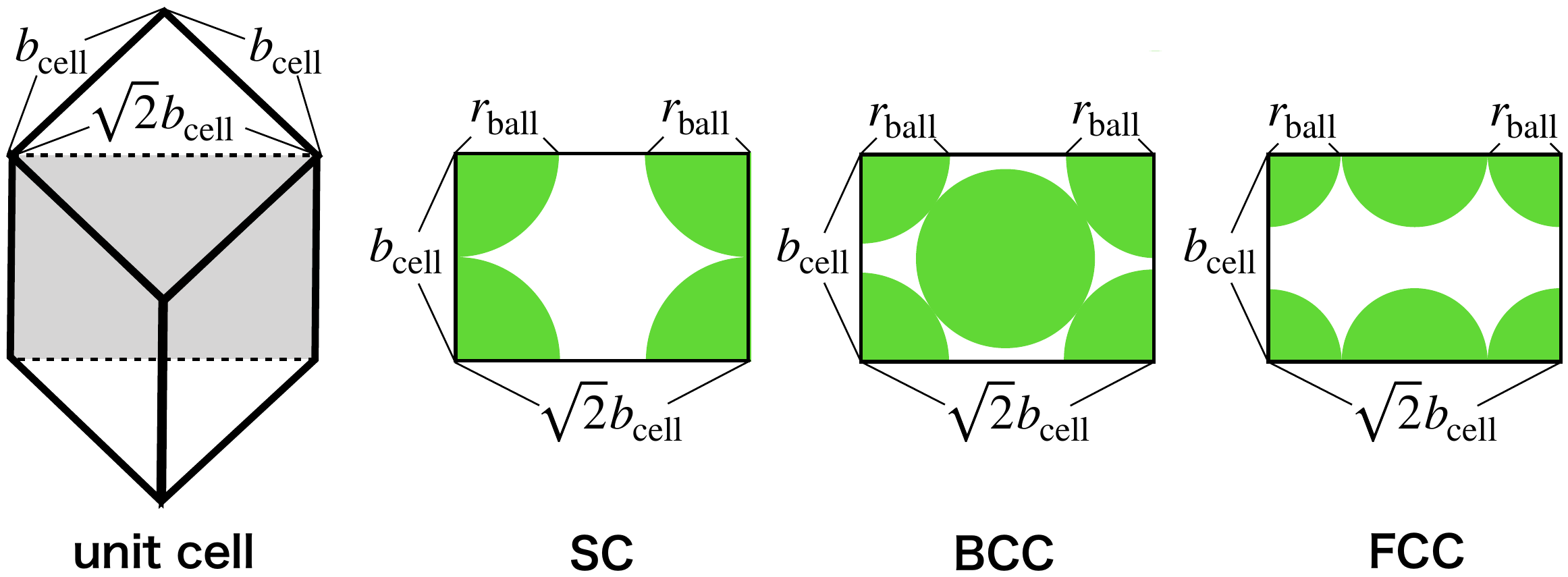}\\
\end{center}
\vspace{-5mm}
\caption{Cross section of the unit cell with SC, BCC, and FCC crystal structures. Green areas indicate the balls.}
\label{fig:crystal}
\end{figure}

{\crd
As an alternative way to reduce the statistical error of the correlator, we develop  a variation of the $ {\mathbb Z_3}$ grid source method \cite{Li:2010pw,Gong:2013vja,Wu:2018tvt}. 
We first divide 
the three-dimensional source space into cubic cells 
with a crystal structure such as simple-cubic (SC) or body-centered-cubic (BCC) or face-centered-cubic (FCC), whose edge length is $b_{\rm cell}=L/n_{\rm div}\in{\mathbb Q}$ with  $n_{\rm div}\in {{\mathbb N}}$ in the lattice unit $a$. Then, we pack equal balls in a nonoverlapping way, where each center of the balls is arranged at each lattice site of the cells. The situation is illustrated in Fig.~\ref{fig:crystal}. 
For the SC, BCC, and FCC structures, 
the radius, the number, and the packing rate of the balls $(r_{\rm ball}, n_{\rm ball}, w_{\rm ball})$ are given by $(b_{\rm cell}/2, n_{\rm div}^3, 52\%)$, $(\sqrt{3}b_{\rm cell}/4, 2n_{\rm div}^3, 68\%)$, and $(\sqrt{2}b_{\rm cell}/4, 4n_{\rm div}^3, 74\%)$, respectively. 
The construction of the crystal source $\Psi(x)$ is completed by assigning ${{\mathbb Z_3}}$ noises $\eta(y)$ to the balls randomly as
\begin{eqnarray}
 \Psi(x) = 
 \sum_{y\in\Lambda_{\rm cell}} \psi(x,y)\, \eta(y),
\label{def_src}
\end{eqnarray}
where $\Lambda_{\rm cell}$ is the set of the lattice sites of the SC/BCC/FCC cells and the function $\psi(x,y)$ has a finite support inside: $\psi(x,y)=1$ for $|x-y|<r_{\rm ball}$.  
Thanks to the ${{\mathbb Z_3}}$ noises, the quark contractions for hadron operators are restricted within each ball effectively.
Note that any smeared wave functions for $\psi(x,y)$ can be introduced inside the balls to maximize the overlapping to the ground state of the target hadron. It is also possible to consider the space filling of polyhedra instead of the ball packing, where the corresponding polyhedron is simple cube, truncated octahedron, and rhombic dodecahedron for the SC, BCC, and FCC cells. 
 
In Sec. \ref{subsec:fse_10b}, to confirm the result for the $\Omega$ baryon correlator obtained with the AMA technique, we repeat the calculation with the BCC crystal source using one ${{\mathbb Z_3}}$ random noise set for each measurement, where we employ the mixed precision nested BiCGStab solver~\cite{pacs-cs1} with the strict tolerance of $|Dx-b|/|b|<10^{-15}$. The parameters for the BCC crystal source are $(n_{\rm div}, b_{\rm cell}, r_{\rm ball}, n_{\rm ball})=(4, 32, 13.9, 128)$ on $128^4$ lattice and (2, 32, 13.9, 16) on $64^4$ lattice. We adopt the exponential smearing function with $A_{\rm s}=1.0$, $B_{\rm s}=0.06$ for $\psi(x,y)$ in Eq.~(\ref{def_src}). The radii of the balls and the smearing parameters are the same on $128^4$ and $64^4$ lattices. 
The number of the measurement is 2560 and 25600 for $128^4$ and $64^4$ lattices, respectively.
}
 
\subsection{Possible mixing states}
\label{subsec:mixing}
\begin{table}[t!]
\caption{Possible mixing states on the lattice. $j$ and $j^\prime$ are the lowest and  the next contributing angular momentum in each octahedral irreducible representation (rep.), respectively.}
\begin{center}
\begin{tabular}{ccccccc}
\hline
Channel & $^2O$ rep. & $j$ & $j^\prime$ & nearby two-body state \\\hline
 $\pi$       & A$_1$ & 0   & 4   & \\
 $K$         & A$_1$ & 0   & 4   & \\
 $\rho$      & G$_1$ & 1   & 3   & $\pi\pi$ \\
 $K^*$       & G$_1$ & 1   & 3   & $K\pi$ \\
 $\phi$      & G$_1$ & 1   & 3   & $KK$ \\
 $N$         & T$_1$ & 1/2 & 7/2 & \\
 $\Lambda$   & T$_1$ & 1/2 & 7/2 & \\
 $\Sigma$    & T$_1$ & 1/2 & 7/2 & \\
 $\Xi$       & T$_1$ & 1/2 & 7/2 & \\
 $\Delta$    & H & 3/2 & 5/2 & $N\pi$\\
 $\Sigma^*$  & H & 3/2 & 5/2 & $\Lambda \pi$\\
 $\Xi^*$     & H & 3/2 & 5/2 & $\Xi \pi$ \\
 $\Omega$    & H & 3/2 & 5/2 & $\Xi K$ \\\hline
\end{tabular}
\end{center}
\label{tab:mixing}
\end{table}%

Since the cubic lattice has only discretized rotational symmetries, it is not possible to assign a definite spin state in the continuum to a single lattice state. Reduction of SU(2) with respect to the double cover of the octahedral group $^2O$ is discussed in detail in Ref.~\cite{grouptheory}. The results are summarized in Table~\ref{tab:mixing}. Spins $j=0$, 1/2, 1, and 3/2 are described by single octahedral irreducible representations A$_1$, G$_1$, T$_1$, and H, respectively. They contain the next contributing angular momenta of $j^\prime=4$, 7/2, 3, and 5/2, and the difference from the ground state is $\Delta j=|j-j^\prime|$=4, 3, 2, and 1, respectively. Note that the $j=3/2$ decuplet state possibly suffers from the contamination from the higher spin state of $j^\prime=5/2$. 

In Table~\ref{tab:mixing}, we also list the nearby two-body state which can mix with the target single-body state. In the real world the $\Omega$ baryon is not allowed to decay into the $\Xi K$ state, whose energy level is slightly above the $\Omega$ baryon mass $(m_\Xi+ m_K)-m_\Omega\approx 140$ MeV. On the lattice, however, both can mix with each other. Since the energy level of the $\Omega$ baryon should be affected by the nearby states belonging to the H representation of the $^2O$ group, it could show an ``unstablelike'' behavior on the lattice similar to those for other decuplet baryons.

\section{Numerical results}
\label{sec:results}

\subsection{Vector meson sector}
\label{subsec:fse_v}

\begin{figure}[h]
\begin{center}
\includegraphics[width=70mm,angle=0]{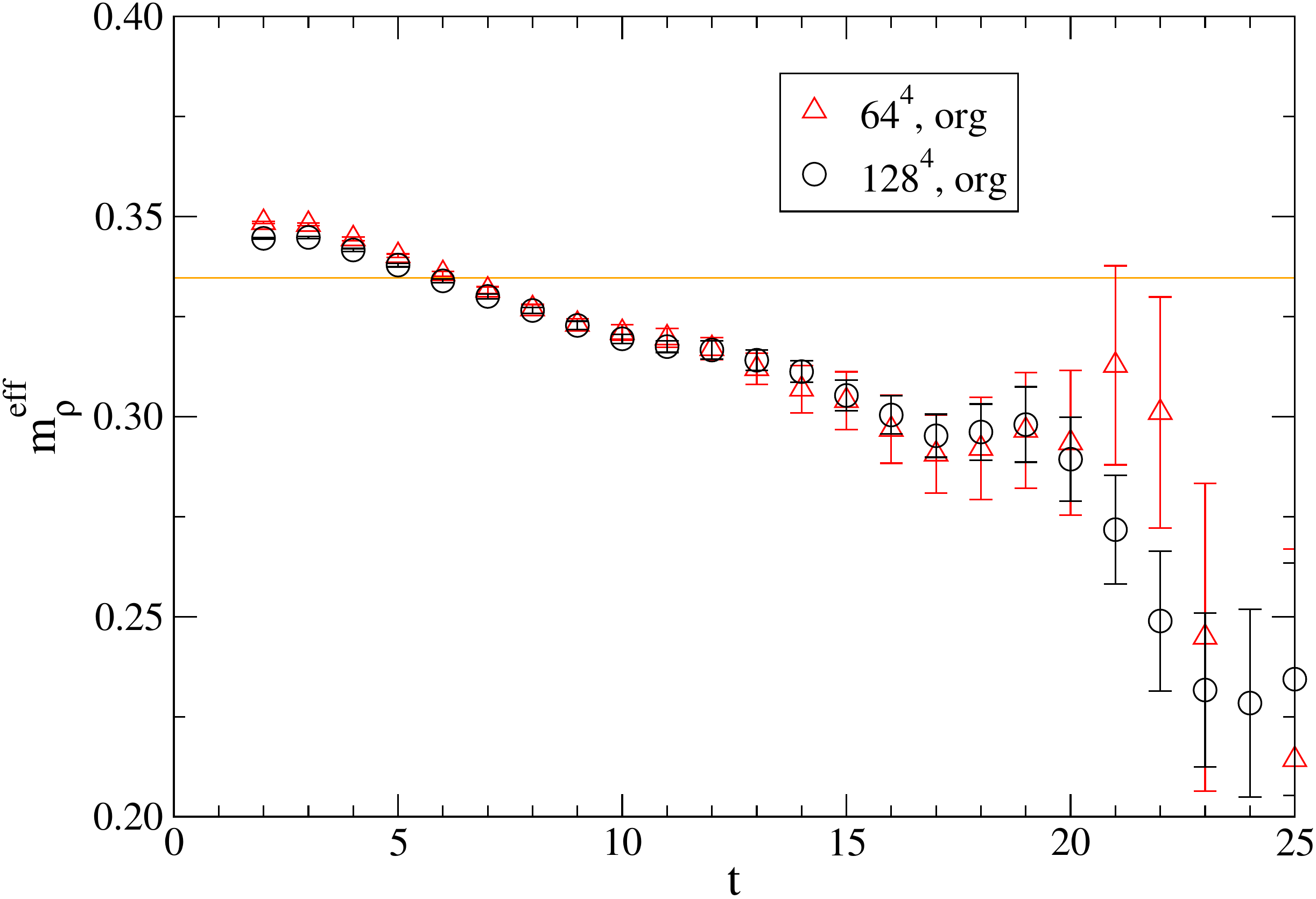}\\
\includegraphics[width=70mm,angle=0]{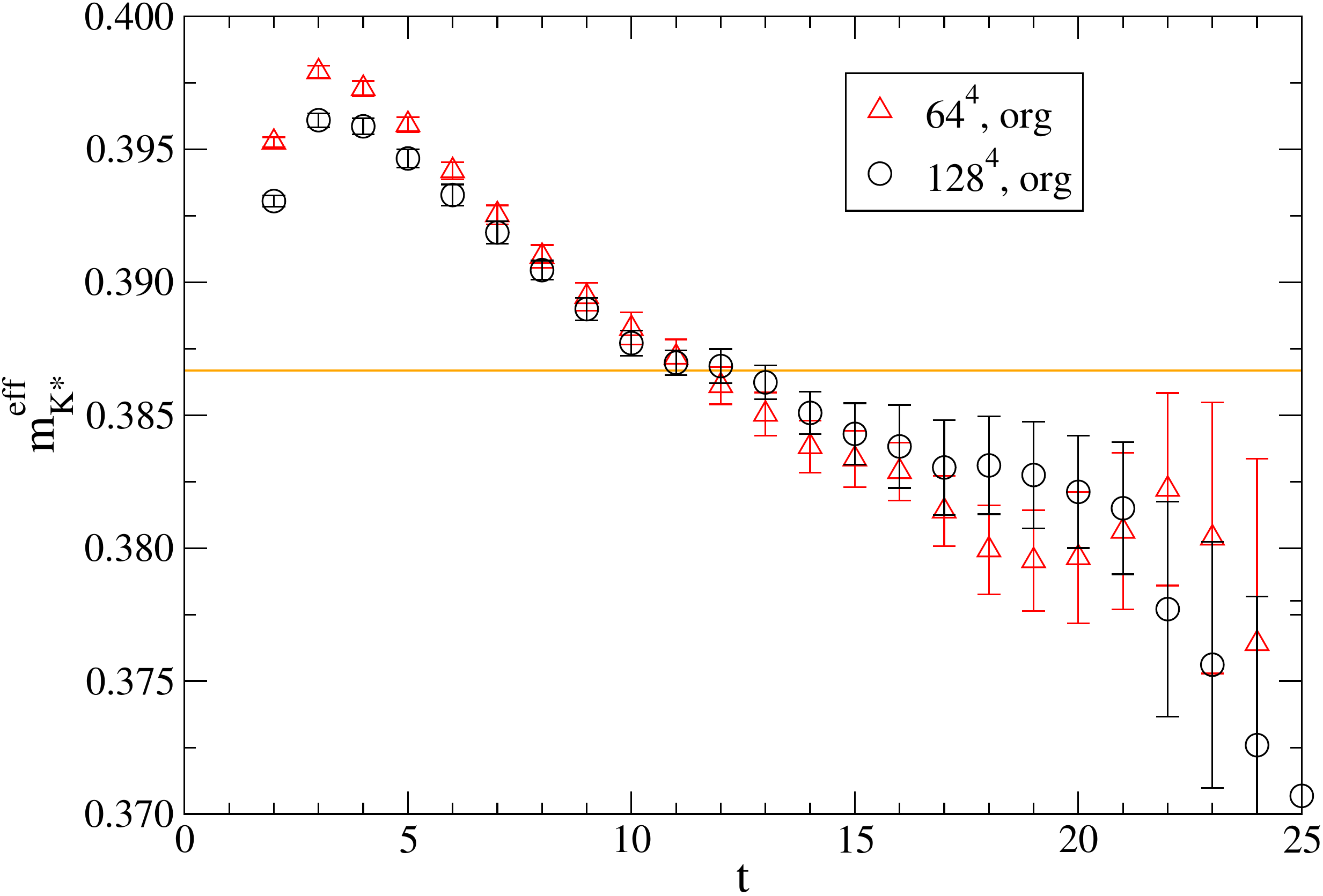}\\
\includegraphics[width=70mm,angle=0]{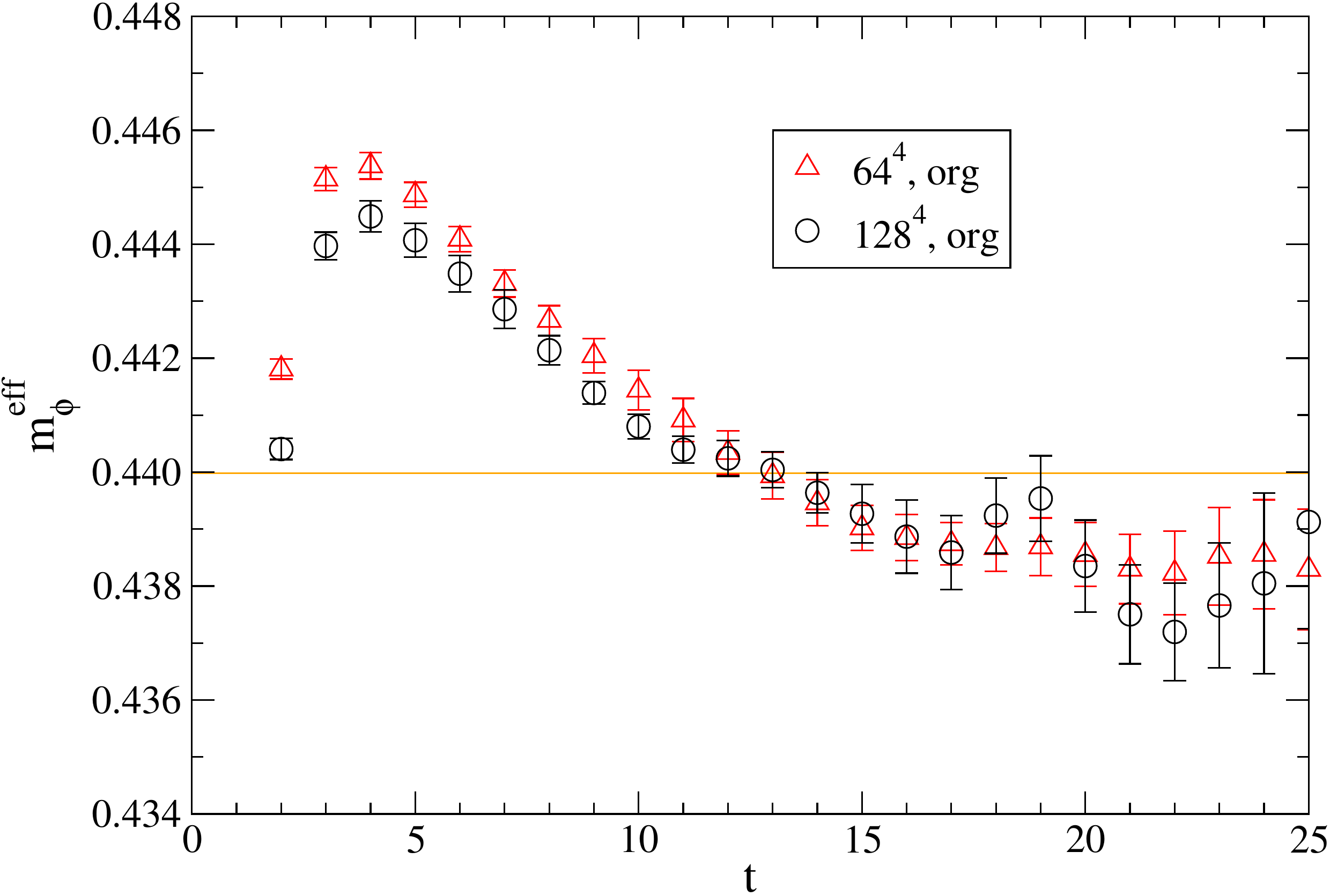}\\
\end{center}
\vspace{-5mm}
\caption{Comparison of effective masses for $\rho$, $K^*$, and $\phi$ (from top to bottom) channels on $64^4$ and $128^4$ lattices. The horizontal line denotes the experimental resonance level in the lattice unit, whose scale is determined by the $\Xi$ baryon mass.}
\label{fig:effmass_v}
\end{figure}

Figure~\ref{fig:effmass_v} shows the effective masses in the vector meson sector.
Red triangle and black circle represent the results on $64^4$ and $128^4$ lattices, respectively, at the original hopping parameters $(\kappa_{\rm ud},\kappa_{\rm s})=(0.126117,0.124902)$. 
We observe that the effective masses for all the  $\rho$, $K^*$, and $\phi$ channels go below the horizontal line in the large time region, which denotes the experimental resonance level with the lattice cutoff determined in Sec.~\ref{sec:physicalpoint}. This could be due to the lower energy level of the $\pi\pi$, $K\pi$, and $KK$ states. We hardly detect the finite size effect beyond the statistical errors. Since it is difficult to find any reasonable plateau for all the channels, we do not try to extract the masses by applying the single exponential fit to the propagators. 

\begin{figure}[h]
\begin{center}
\includegraphics[width=70mm,angle=0]{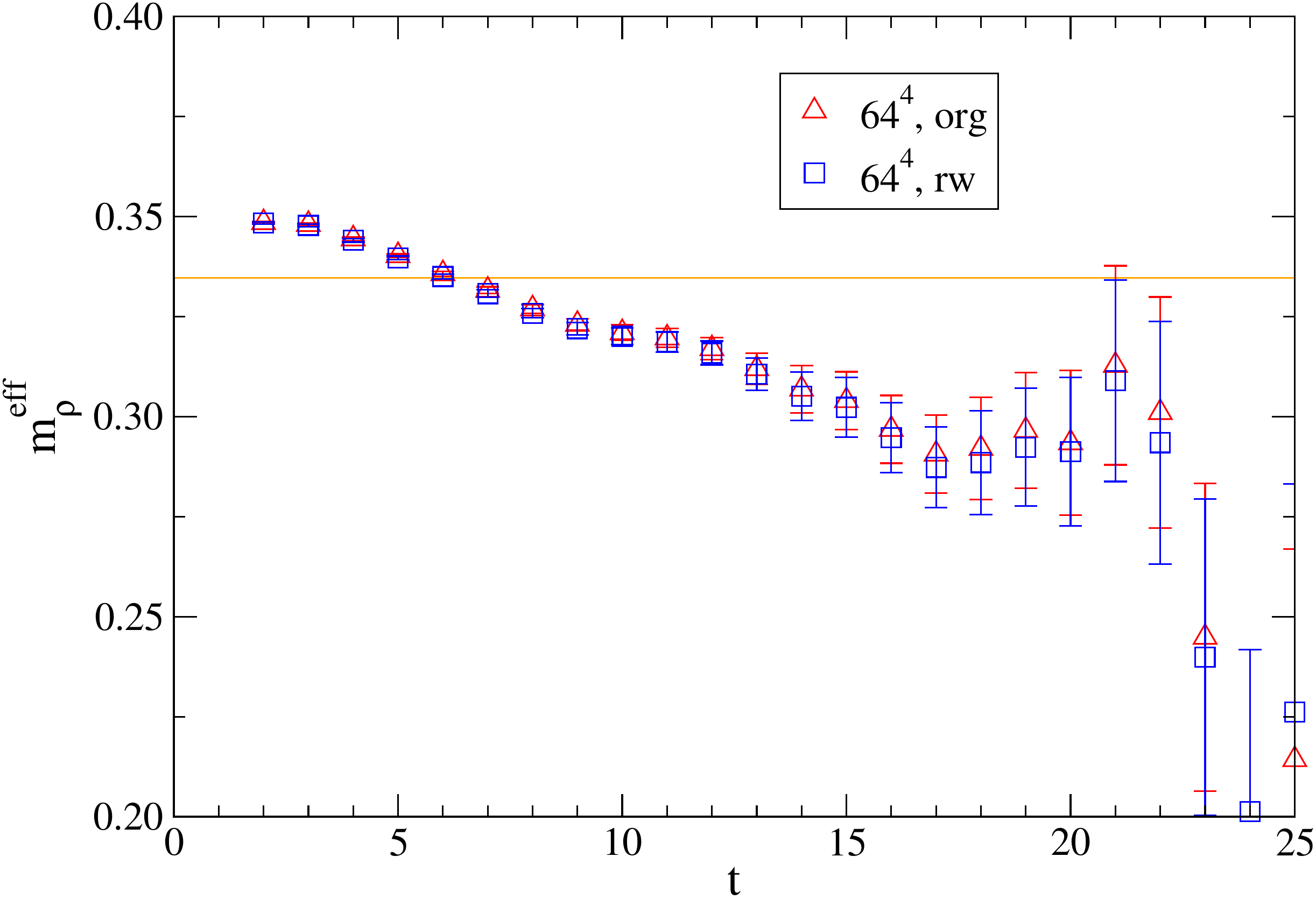}\\
\includegraphics[width=70mm,angle=0]{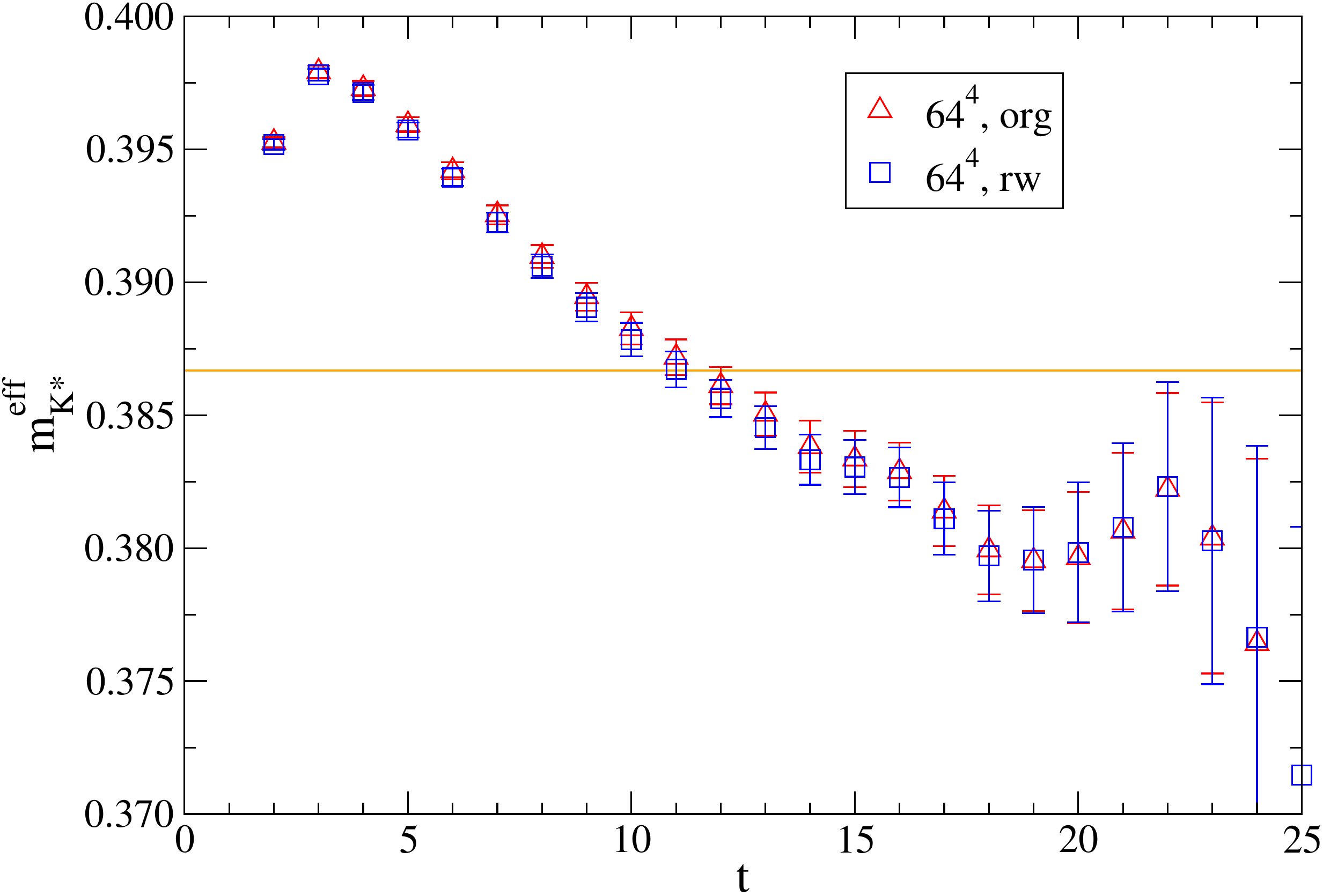}\\
\includegraphics[width=70mm,angle=0]{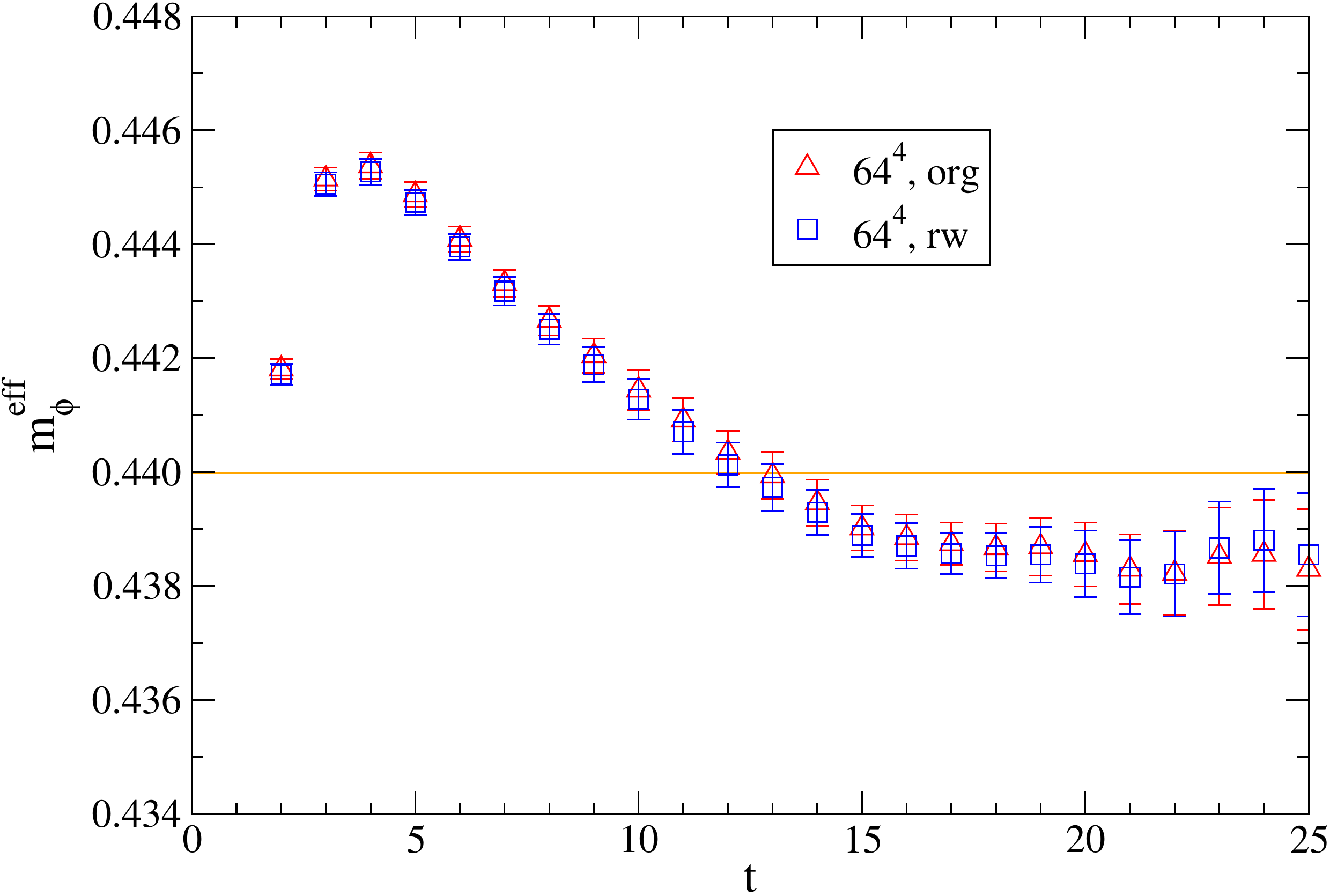}\\
\end{center}
\vspace{-5mm}
\caption{Comparison of effective masses for $\rho$, $K^*$, and $\phi$ (from top to bottom) channels on the $64^4$ lattice at the original hopping parameters and the reweighted ones $(\kappa_{\rm ud}^*,\kappa_{\rm s}^*)=(0.126119,0.124902)$. The horizontal line is the same as in Fig.~\ref{fig:effmass_v}.}
\label{fig:effmass_v_64}
\end{figure}

We also make another comparison by adjusting the AWI ud quark masses on the $64^4$ lattice to those on the $128^4$ lattice with the use of the reweighting technique employed in Ref.~\cite{mf_nf2+1}.  In Fig.~\ref{fig:effmass_v_64}, we compare the effective masses on the $64^4$ lattice at the original hopping parameters and the reweighted ones $(\kappa_{\rm ud}^*,\kappa_{\rm s}^*)=(0.126119,0.124902)$, where the light and strange quark masses are consistent with those at the original hopping parameters on the $128^4$ lattice. We find that both data are almost degenerate so that the reweighting effect is negligible within the current statistical errors. So we can conclude that the effective masses in the vector meson sector does not show the finite size effect beyond the statistical errors in both cases of fixing $\kappa$ values or measured AWI quark masses.

\subsection{Octet baryon sector}
\label{subsec:fse_8b}

\begin{figure}[t!]
\begin{center}
\includegraphics[width=70mm,angle=0]{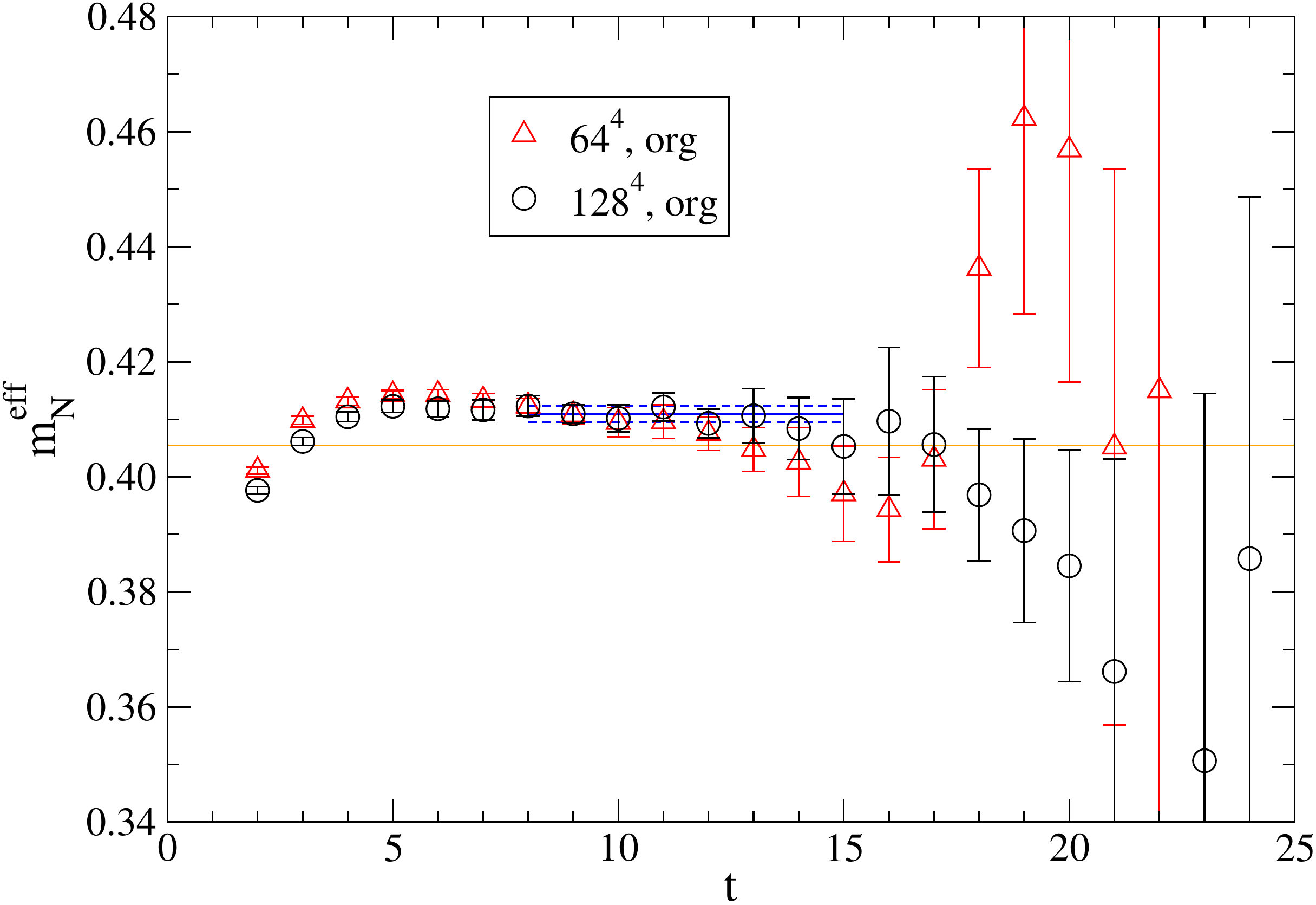}\\
\includegraphics[width=70mm,angle=0]{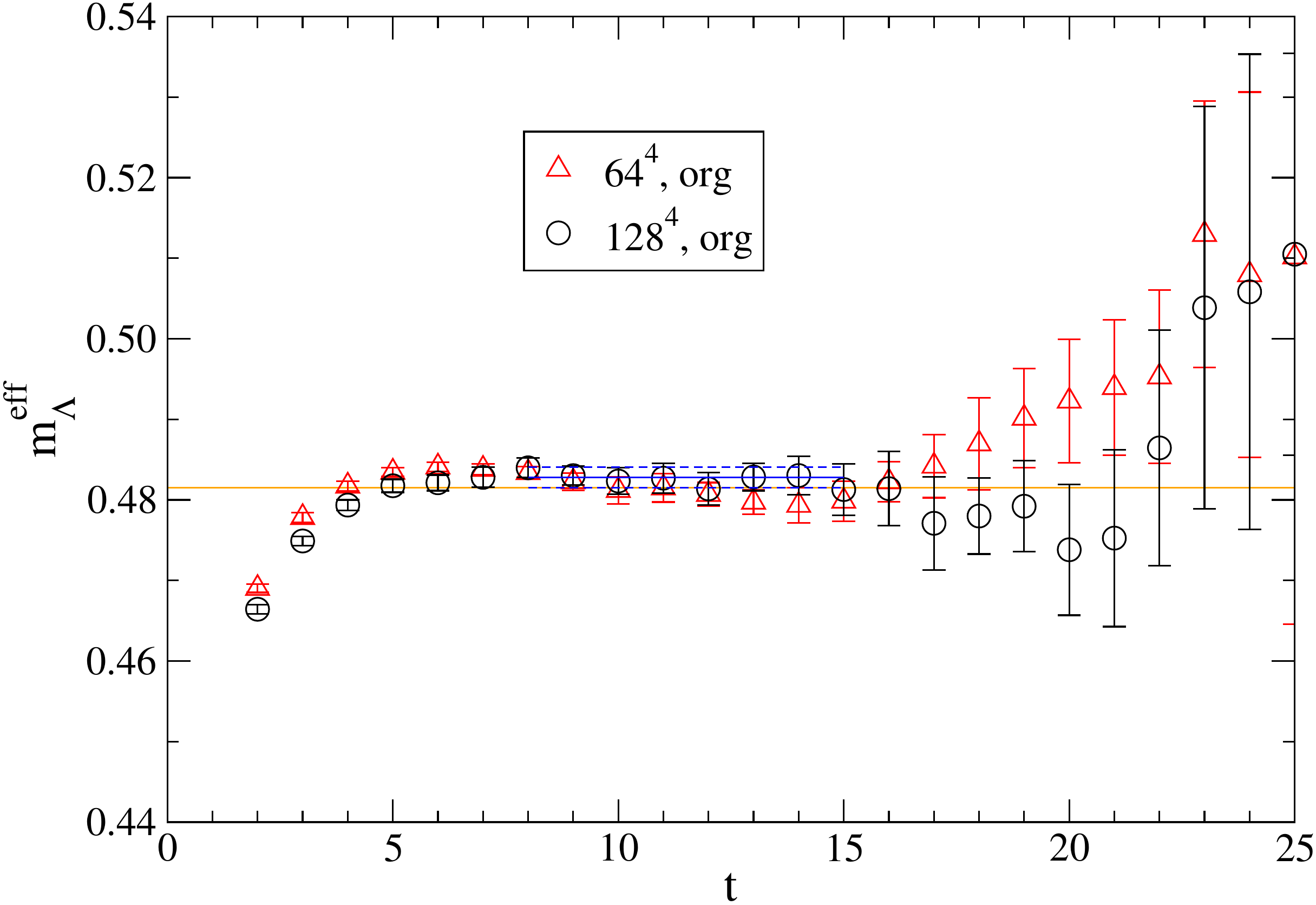}\\
\includegraphics[width=70mm,angle=0]{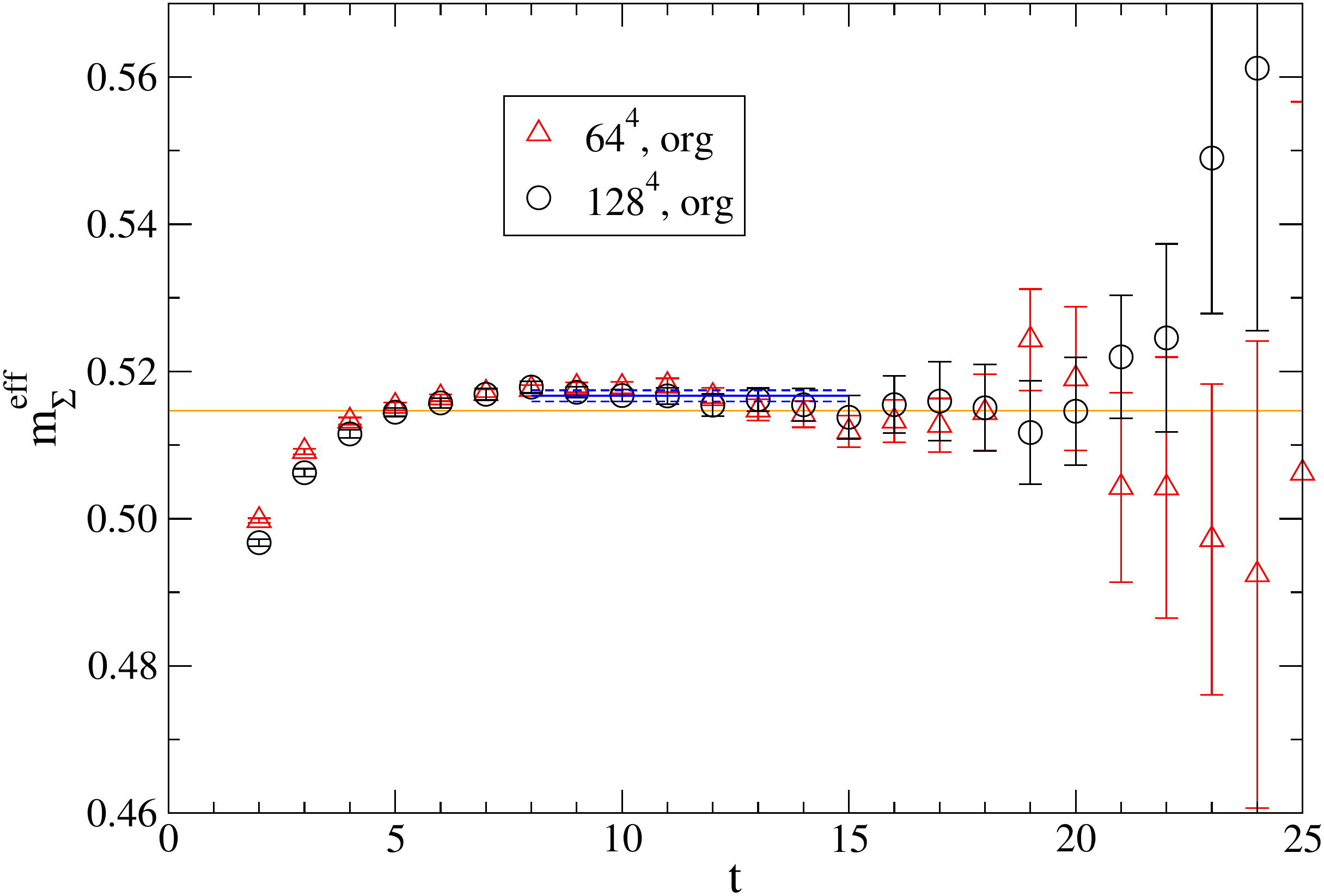}\\
\includegraphics[width=70mm,angle=0]{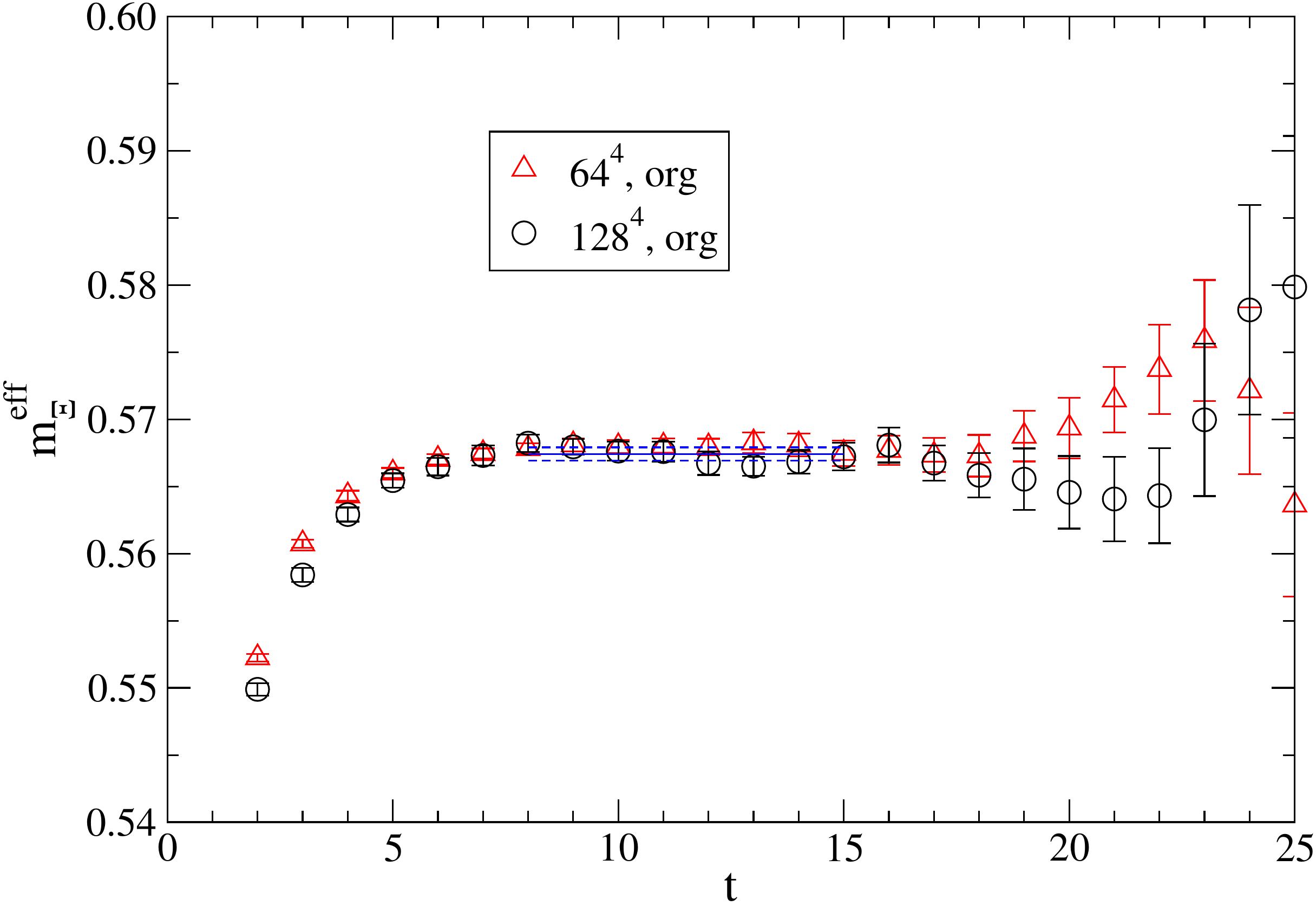}\\
\end{center}
\vspace{-5mm}
\caption{Comparison of effective masses for $N$, $\Lambda$, $\Sigma$, and $\Xi$ (from top to bottom) baryon channels on $64^4$ and $128^4$ lattices. The horizontal line denotes the experimental mass in the lattice unit.}
\label{fig:effmass_8b}
\end{figure}

The effective masses in the octet baryon sector are plotted in Fig.~\ref{fig:effmass_8b}.
Red triangle and black circle  represent the results on $64^4$ and $128^4$ lattices, respectively, at the original hopping parameters. 
The horizontal line denotes the experimental value with the lattice cutoff determined in Sec.~\ref{sec:physicalpoint}.
It is hard to detect the finite size effect for all the $N$, $\Lambda$, $\Sigma$, and $\Xi$ channels. This is also true in the comparison with the reweighted results on $64^4$ lattice as for the vector meson channels. We observe that the effective mass for the $\Xi$ channel shows a good plateau with the smallest statistical errors among the octet baryons so that we employ it together with the $\pi$ and $K$ meson masses to determine the physical quark masses and the cutoff scale in Sec.~\ref{sec:physicalpoint}.
In Table~\ref{tab:fitmass_8b}, we summarize the fit results for the octet baryon masses choosing the fit range of $[t_{\rm min},t_{\rm max}]=[8,15]$ on $128^4$ and $64^4$ lattices. We observe that the results at the original hopping parameters on both lattices are consistent within the statistical errors, which are 0.5\% level for the nucleon and down to 0.1\% level for the $\Xi$ baryon.

\begin{table}[t!]
\caption{Fit results for octet baryon masses on $128^4$ and $64^4$ lattices at the original hopping parameters and the reweighted ones $(\kappa_{\rm ud}^*,\kappa_{\rm s}^*)=(0.126119,0.124902)$.}
\begin{center}
\begin{tabular}{ccccccc}
\hline
Mass        & $128^4$ (org)     & $64^4$ (org) & $64^4$ (rw) \\\hline
$m_{N}$       & 0.4109(14)  & 0.4094(21)   & 0.4074(26)\\
$m_{\Lambda}$ & 0.4828(13)  & 0.4815(12)   & 0.4803(16)\\
$m_{\Sigma}$  & 0.51669(75) & 0.51720(78)  & 0.51647(86) \\
$m_{\Xi}$     & 0.56742(50) & 0.56802(46)  & 0.56777(49)\\\hline
\end{tabular}
\end{center}
\label{tab:fitmass_8b}
\end{table}%

\begin{figure}[t!]
\begin{center}
\includegraphics[width=70mm,angle=0]{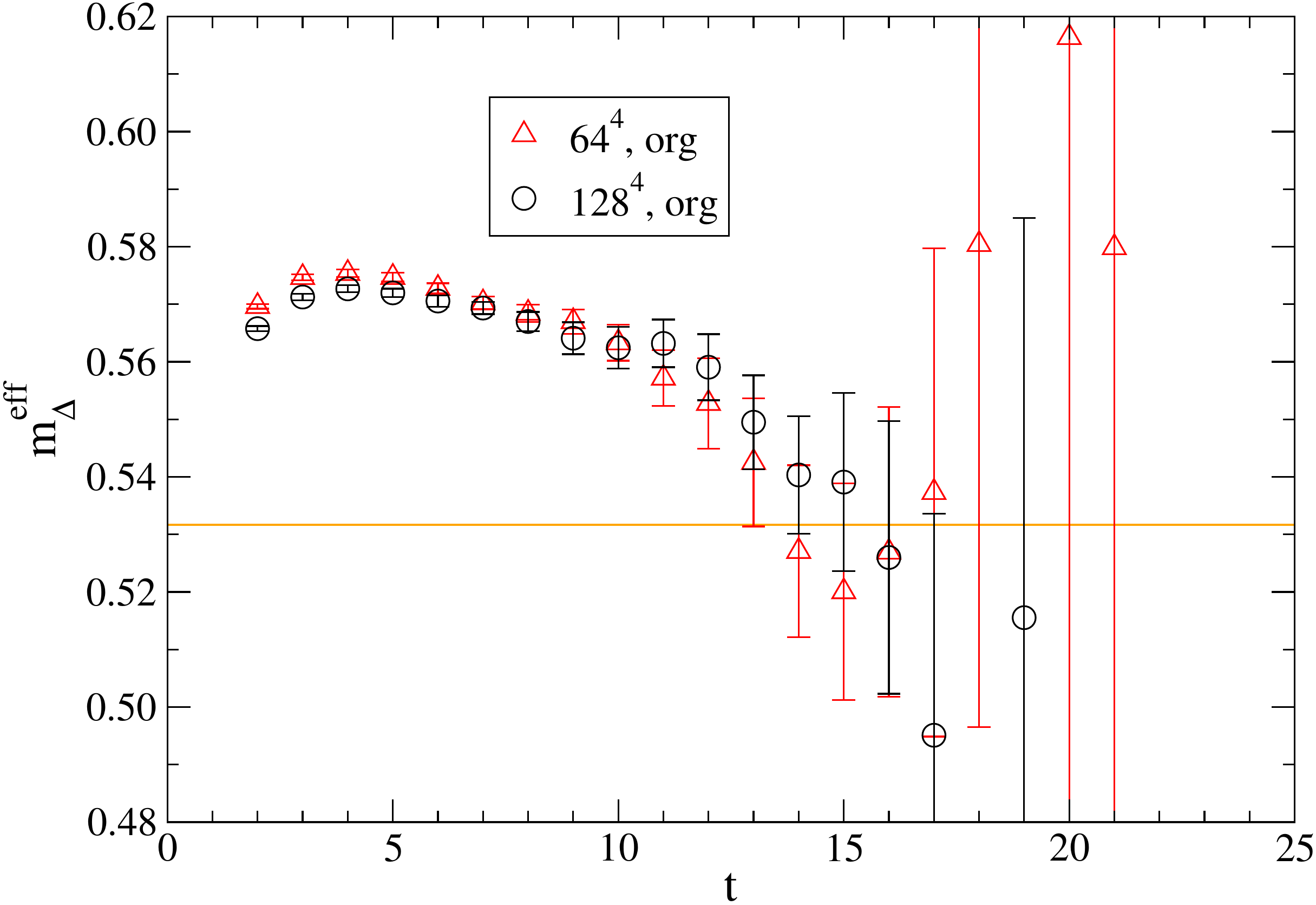}\\
\includegraphics[width=70mm,angle=0]{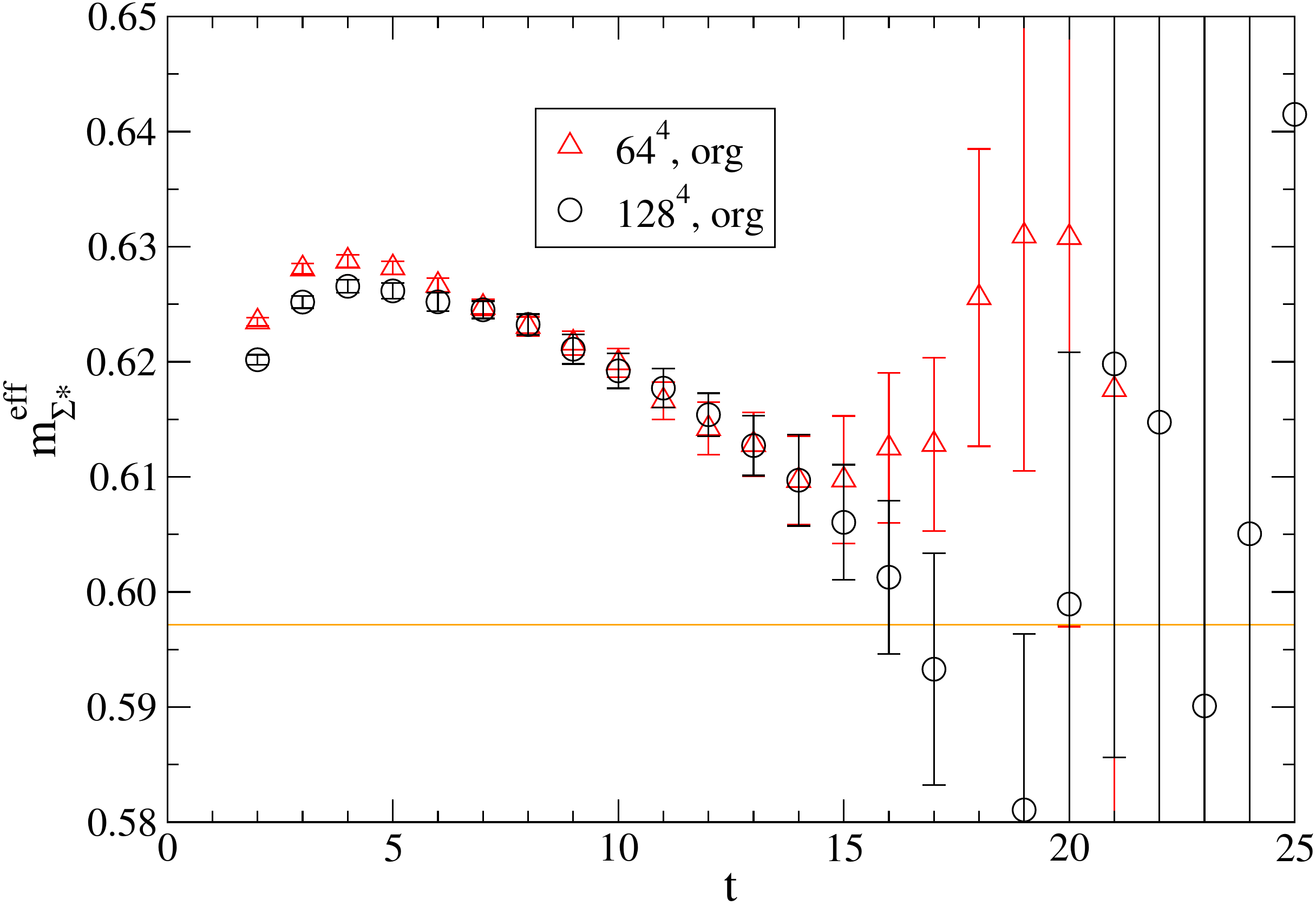}\\
\includegraphics[width=70mm,angle=0]{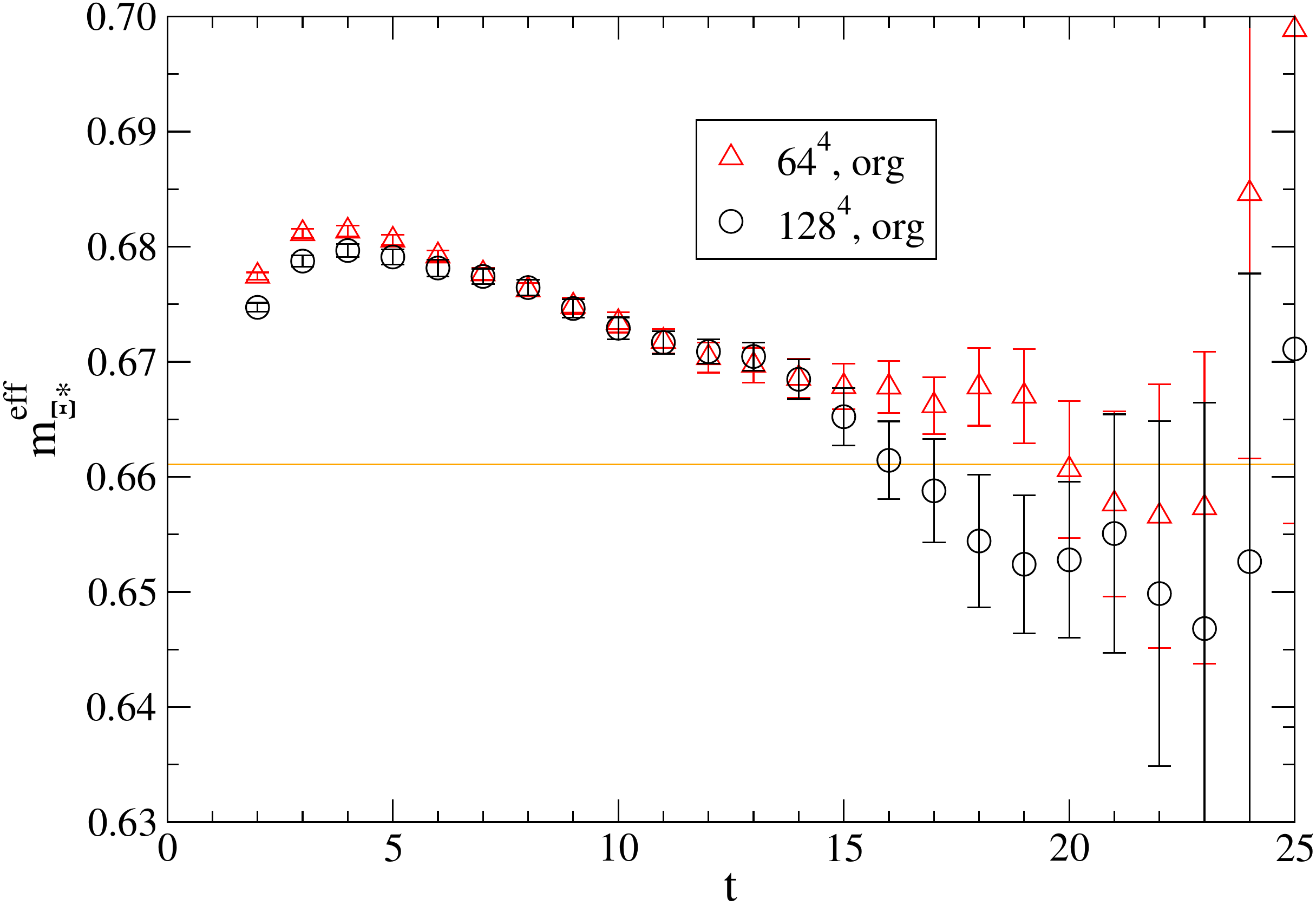}\\
\includegraphics[width=70mm,angle=0]{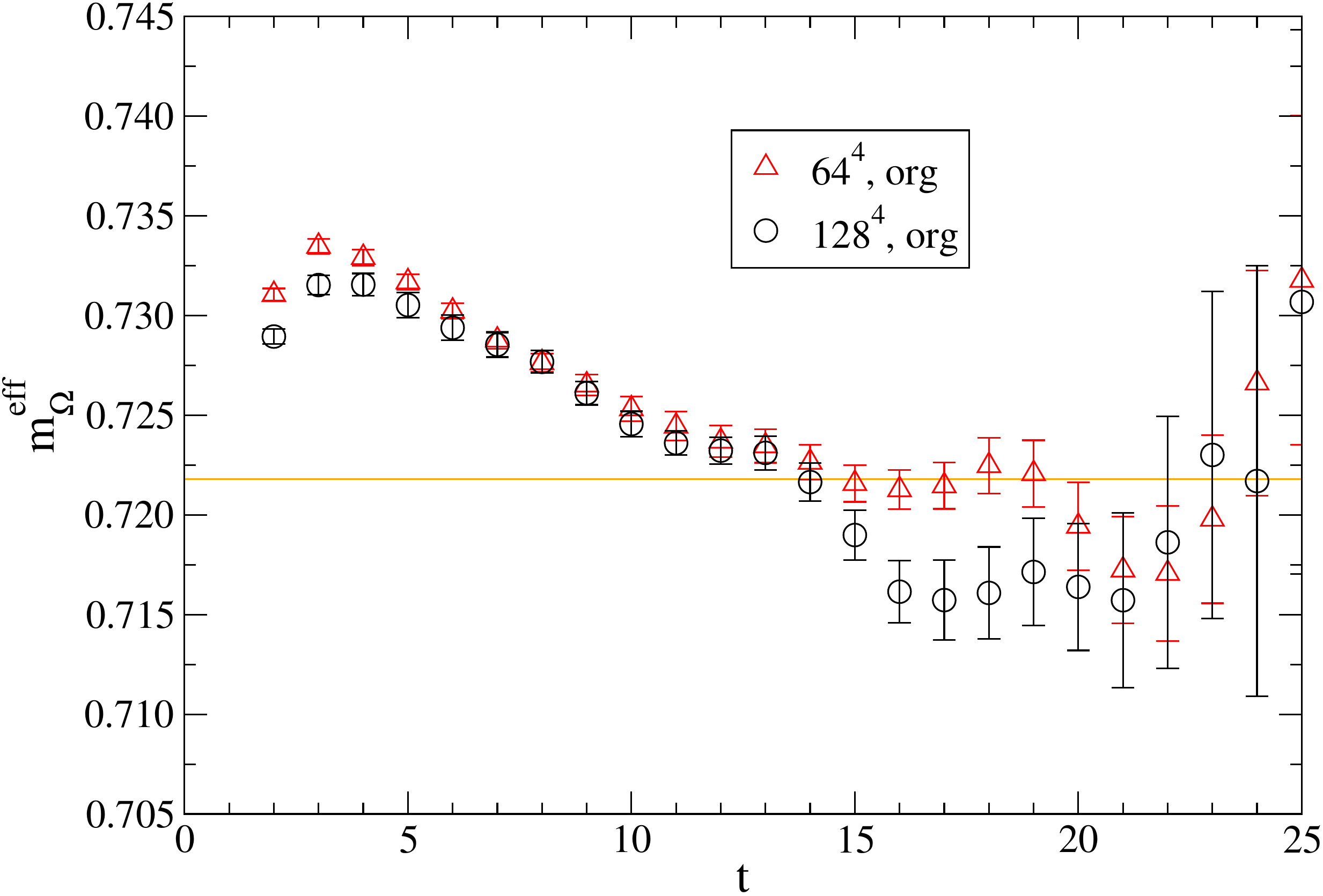}\\
\end{center}
\vspace{-5mm}
\caption{Comparison of effective masses for $\Delta$, $\Sigma^*$, $\Xi^*$, and $\Omega$ (from top to bottom) baryon channels on $64^4$ and $128^4$ lattices. The horizontal line denotes the experimental resonance level for $\Delta$, $\Sigma^*$, and $\Xi^*$ and the experimental mass for $\Omega$.}
\label{fig:effmass_10b}
\end{figure}

\subsection{Decuplet baryon sector}
\label{subsec:fse_10b}

\begin{figure}[t!]
\begin{center}
\includegraphics[width=70mm,angle=0]{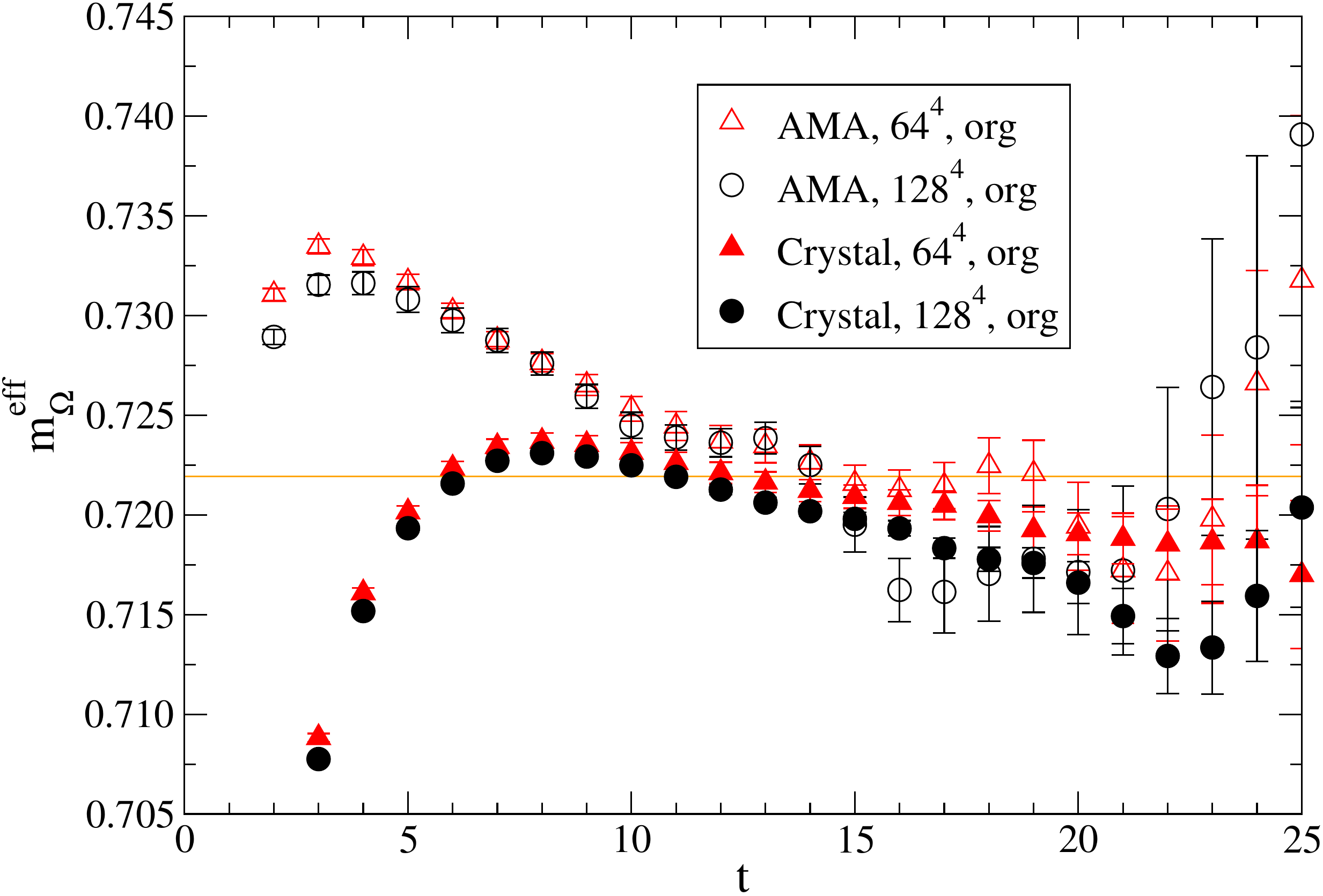}\\
\end{center}
\vspace{-5mm}
\caption{Comparison of effective mass for $\Omega$ baryon channel on $64^4$ and $128^4$ lattices. Open symbols are the same as in the bottom panel of Fig.~\ref{fig:effmass_10b}. The horizontal line denotes the experimental $\Omega$ baryon mass.}
\label{fig:effmass_omega}
\end{figure}

We plot the effective masses for the $\Delta$, $\Sigma^*$, $\Xi^*$, and $\Omega$ channels in Fig.~\ref{fig:effmass_10b}. Red triangle and black circle represent the results on $64^4$ and $128^4$ lattices, respectively, at the original hopping parameters. 
The horizontal lines denote the experimental values of the resonance levels for the $\Delta$, $\Sigma^*$, and $\Xi^*$ baryons with the lattice cutoff determined in Sec.~\ref{sec:physicalpoint}, and it represents the experimental mass for the $\Omega$ baryon. 
The effective masses for the $\Delta$, $\Sigma^*$, $\Xi^*$, and $\Omega$ channels share quite similar features with the vector meson sector: we cannot find any reasonable plateau region to validate the single exponential fit.  It should be also noted that as in the cases of the vector mesons and the octet baryons the results at the original and reweighted hopping parameters on $64^4$ lattice do not show any significant difference. It seems difficult to detect the finite size effect beyond the current statistical errors. But we point out a possibility that the effective mass for the $\Omega$ channel may indicate a finite size effect between $64^4$ and $128^4$ lattices at $t\simgt 15$. In order to check whether this signal is caused by a finite size effect or not, we repeat the calculation of the $\Omega$ propagator with a higher level of precision  employing  the BCC crystal source explained in Sec.~\ref{sec:crystal_source} without the AMA technique. Solid red triangles and solid black circles in Fig.~\ref{fig:effmass_omega} represent the results on $64^4$ and $128^4$ lattices at the simulation point, respectively. We observe that there are small discrepancies between the results on $64^4$ and $128^4$ lattices with drastically reduced statistical errors. A more important point is that both results show monotonic decreases toward the large time region, which is quite similar to other decuplet baryon channels. This ``unstablelike'' behavior could be caused by the mixing effect with the nearby states belonging to the H representation of the $^2O$ group.

\section{Determination of physical quark masses and cutoff scale}
\label{sec:physicalpoint}

\begin{figure}[t!]
\begin{center}
\includegraphics[width=70mm,angle=0]{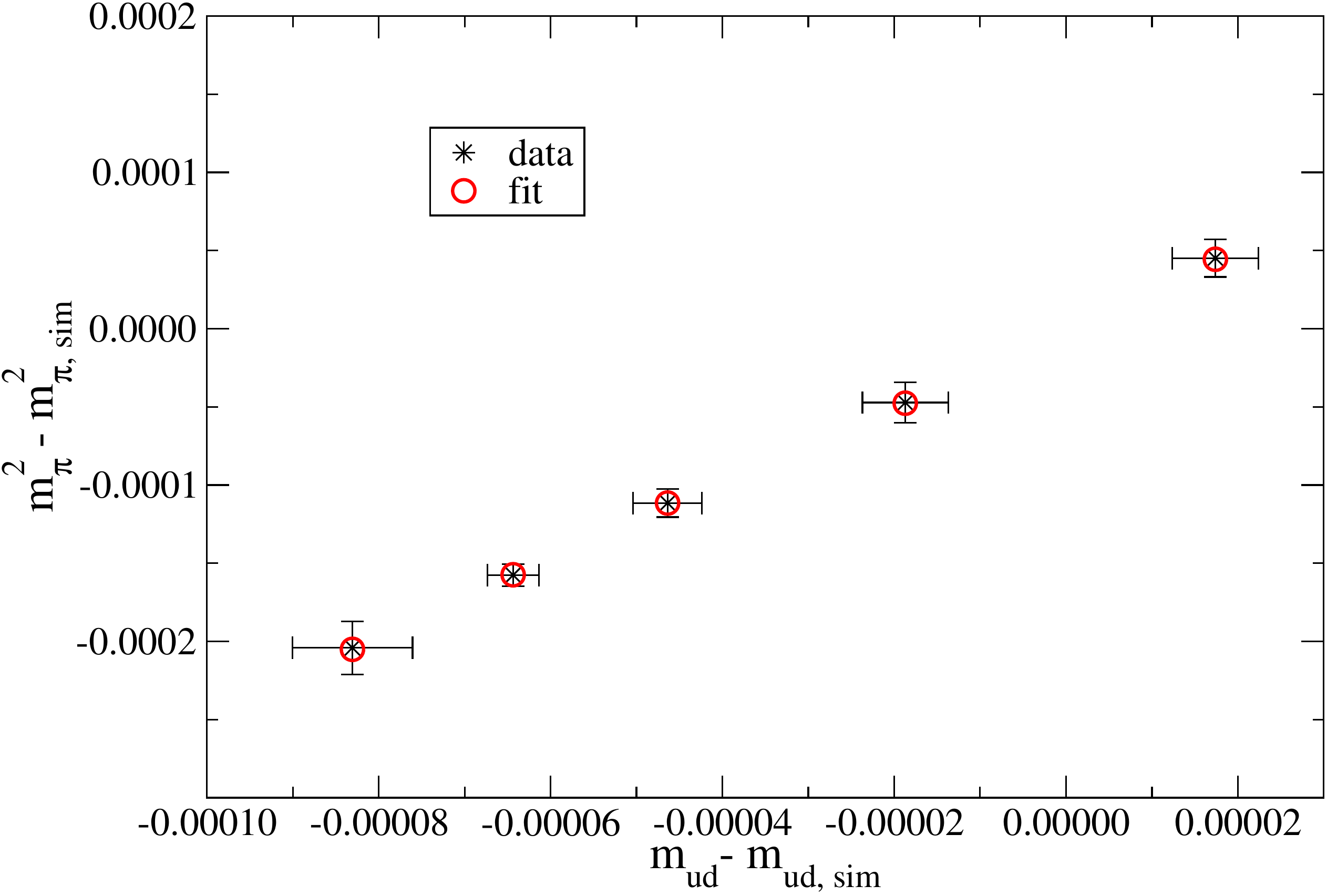}\\
\includegraphics[width=70mm,angle=0]{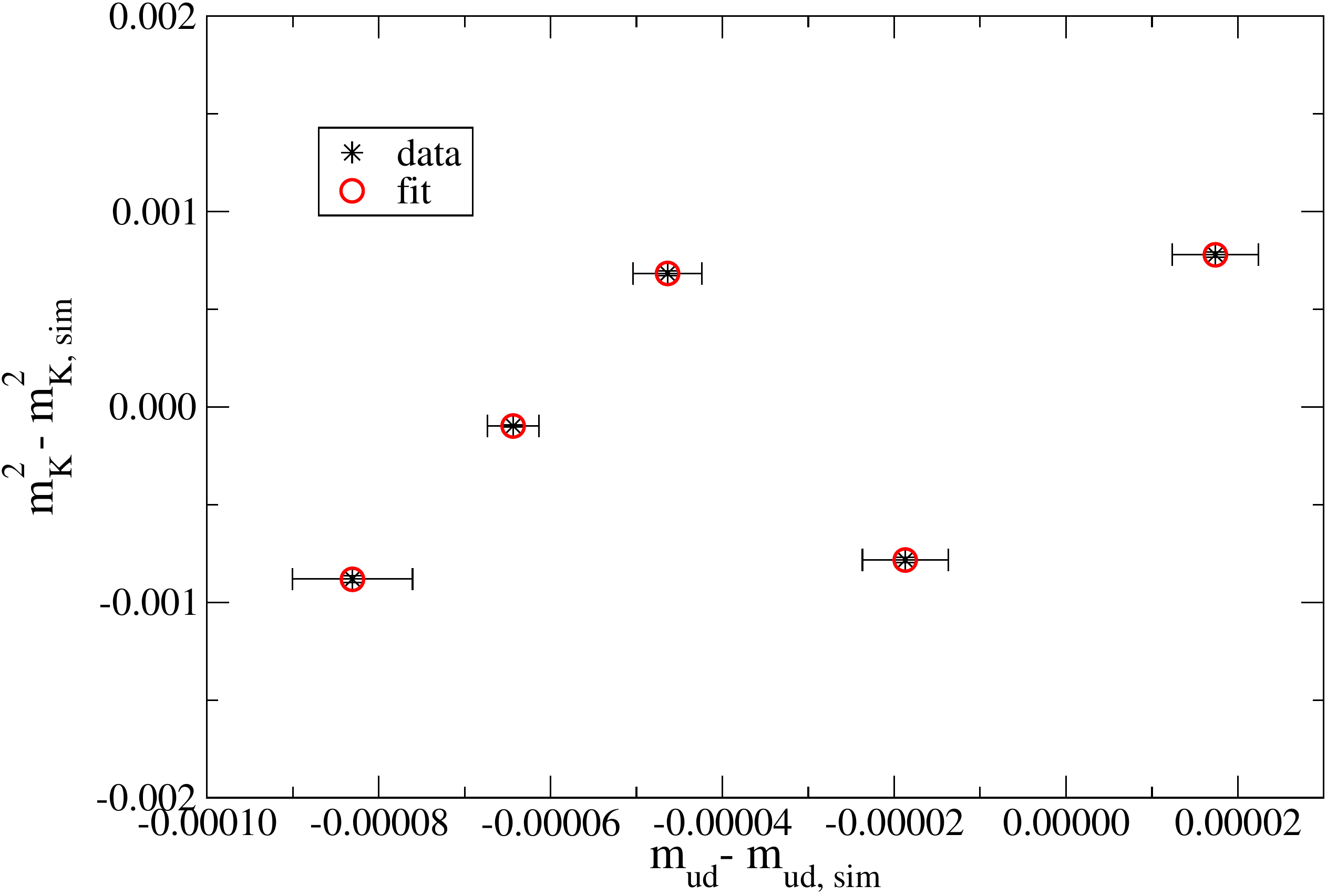}\\
\includegraphics[width=70mm,angle=0]{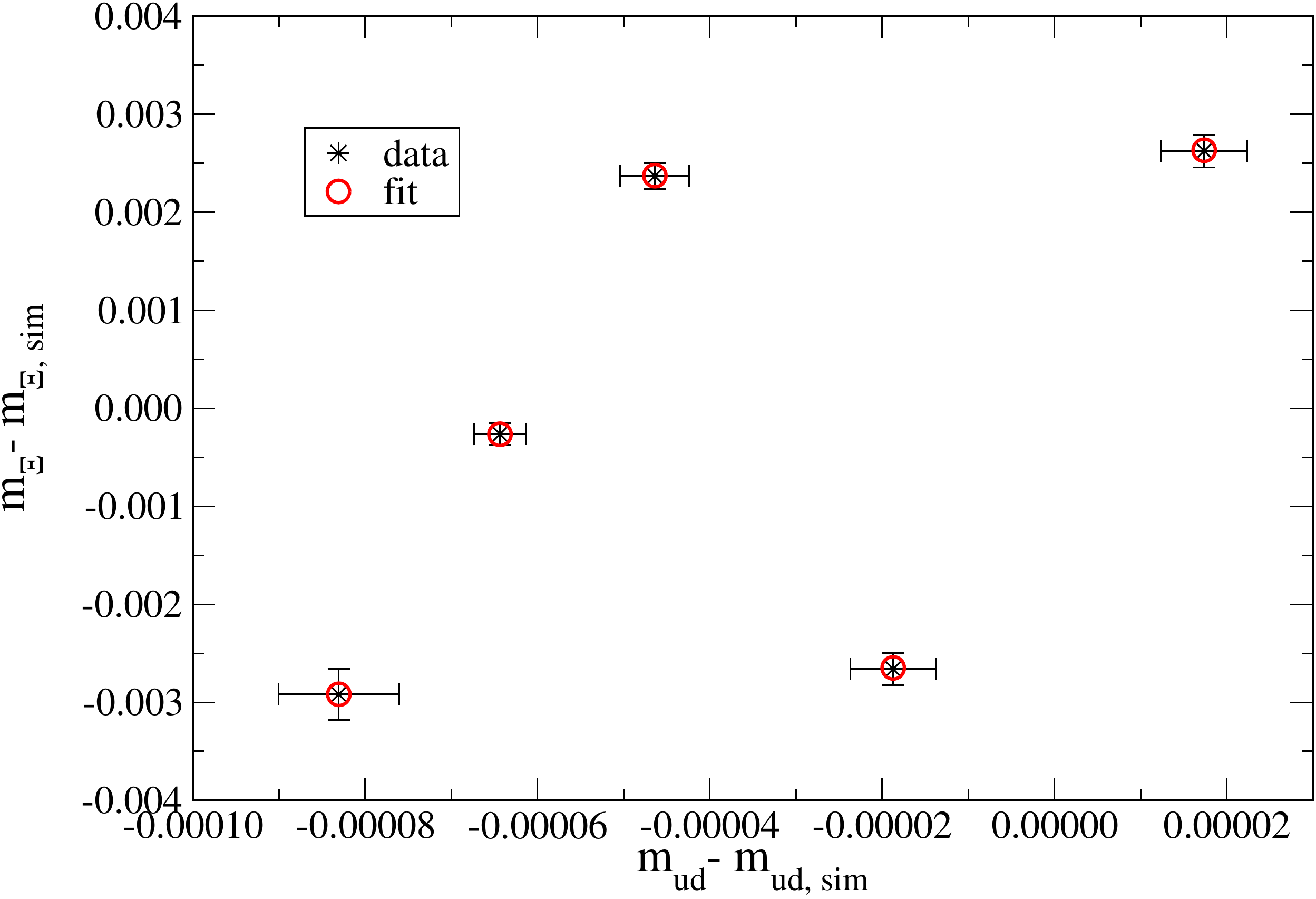}\\
\end{center}
\vspace{-5mm}
\caption{AWI quark mass dependence of $m_\pi^2-m_\pi^2|_{\rm sim}$, $m_K^2-m_K^2|_{\rm sim}$ and $m_\Xi-m_\Xi|_{\rm sim}$ (from top to bottom) on $64^4$ lattice together with the fit results.}
\label{fig:fit_64}
\end{figure}

\begin{figure}[t!]
\begin{center}
\includegraphics[width=75mm,angle=0]{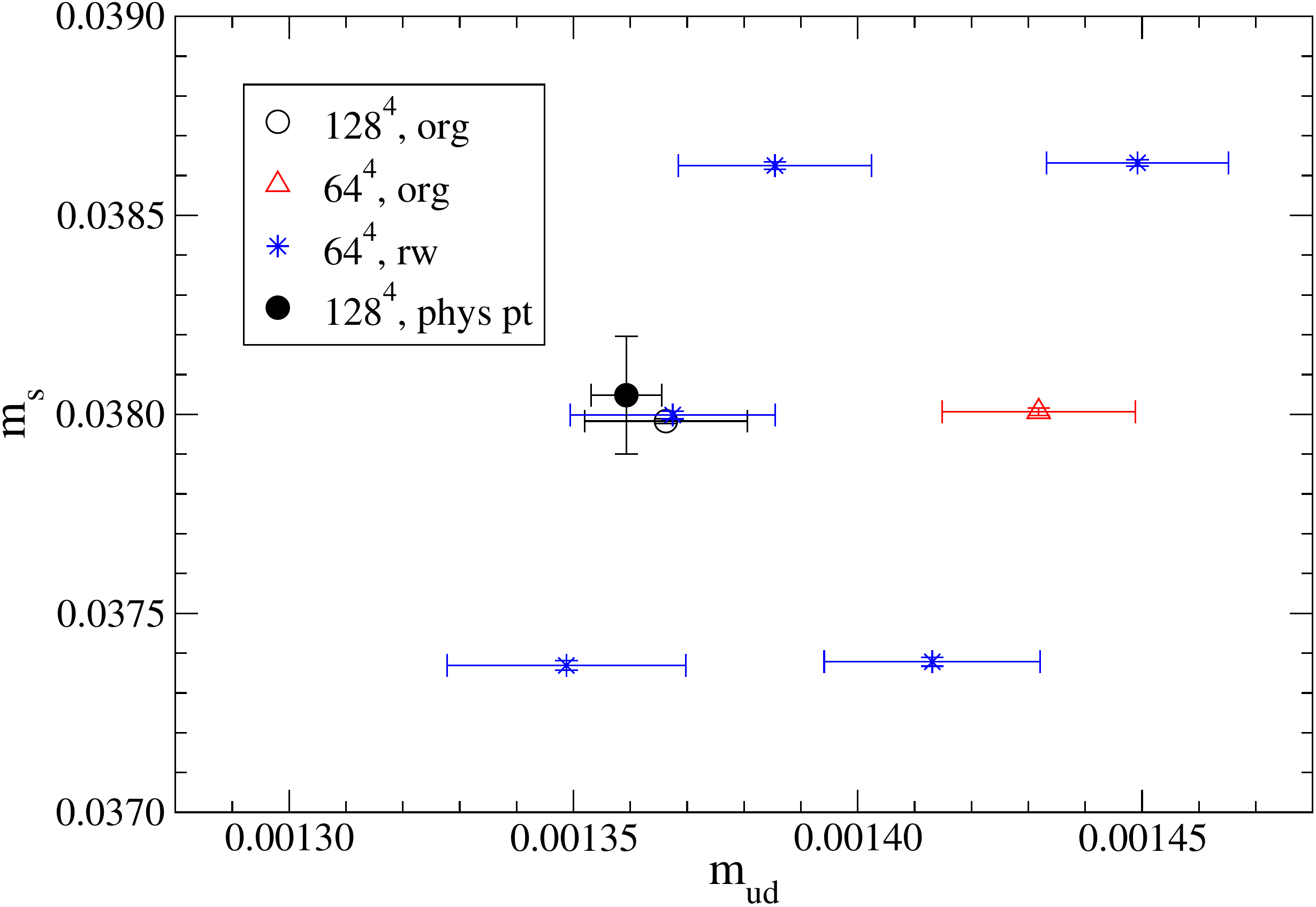}\\
\end{center}
\vspace{-5mm}
\caption{Physical point in $m_{\rm ud}$-$m_{\rm s}$ plane. The light and strange quark masses at the reweighted hopping parameters $(\kappa_{\rm ud}^*,\kappa_{\rm s}^*)=(0.126119,0.124902)$ on $64^4$ lattice are consistent with those at the original hopping parameters on $128^4$ lattice.}
\label{fig:physicalpoint}
\end{figure}

The physical point on the $128^4$ lattice is determined by the Taylor expansion  of $m_\pi^2$, $m_K^2$, and $m_\Xi$ in terms of the AWI quark masses around the simulation point,
\ben
m_\pi^2&=&m_\pi^2\vert_{\rm sim}+\left.\frac{\partial m_\pi^2}{\partial m_{\rm ud}}\right\vert_{\rm sim}\left(m_{\rm ud}- m_{\rm ud}\vert_{\rm sim}\right)\nn\\
&&\hspace{15mm}+\left.\frac{\partial m_\pi^2}{\partial m_{\rm s}}\right\vert_{\rm sim}\left(m_{\rm s}- m_{\rm s}\vert_{\rm sim}\right), \\
m_K^2&=&m_K^2\vert_{\rm sim}+\left.\frac{\partial m_K^2}{\partial m_{\rm ud}}\right\vert_{\rm sim}\left(m_{\rm ud}- m_{\rm ud}\vert_{\rm sim}\right)\nn\\
&&\hspace{15mm}+\left.\frac{\partial m_K^2}{\partial m_{\rm s}}\right\vert_{\rm sim}\left(m_{\rm s}- m_{\rm s}\vert_{\rm sim}\right), \\
m_\Xi&=&m_\Xi\vert_{\rm sim}+\left.\frac{\partial m_\Xi}{\partial m_{\rm ud}}\right\vert_{\rm sim}\left(m_{\rm ud}- m_{\rm ud}\vert_{\rm sim}\right)\nn\\
&&\hspace{15mm}+\left.\frac{\partial m_\Xi}{\partial m_{\rm s}}\right\vert_{\rm sim}\left(m_{\rm s}- m_{\rm s}\vert_{\rm sim}\right). 
\een
The six coefficients $\partial m_\pi^2/\partial m_{\rm ud}$, $\partial m_\pi^2/\partial m_{\rm s}$, $\partial m_K^2/\partial m_{\rm ud}$, $\partial m_K^2/\partial m_{\rm s}$, $\partial m_\Xi/\partial m_{\rm ud}$, and $\partial m_\Xi/\partial m_{\rm s}$ are determined from the simulated and reweighted results on $64^4$ lattice. In Fig.~\ref{fig:fit_64}, we plot the AWI quark mass dependence of $m_\pi^2-m_\pi^2|_{\rm sim}$, $m_K^2-m_K^2|_{\rm sim}$, and $m_\Xi-m_\Xi|_{\rm sim}$ on $64^4$ lattice together with the uncorrelated fit results employing a function of $C_{\rm ud}\cdot \left(m_{\rm ud}- m_{\rm ud}\vert_{\rm sim}\right)$+$C_{\rm s}\cdot\left(m_{\rm s}- m_{\rm s}\vert_{\rm sim}\right)$. We find that the quark mass dependence is well described by the linear function. Using the values of $C_{\rm ud}$ and $C_{\rm s}$ listed in Table~\ref{tab:fit_64}, we determine the physical point on the $128^4$ lattice to reproduce the experimental values of $m_\pi=0.1350$ GeV, $m_K=0.4976$ GeV, and $m_\Xi=1.3148$ GeV~\footnote{The quark mass differences between the simulation point and the physical point are so tiny that we can completely neglect the finite size effects in $C_{\rm ud}$ and $C_{\rm s}$.}. The results for the bare ud and s quark masses and the lattice cutoff are presented in Table~\ref{tab:physicalpoint}. We also plot the physical quark masses in the $m_{\rm ud}$-$m_{\rm s}$ plane of Fig.~\ref{fig:physicalpoint}, which are consistent with the original quark masses at the simulation point. It is confirmed that the original hopping parameters are successfully tuned at the physical point. The cutoff scale we determine in this paper,
\begin{equation}
  a^{-1}=2.3162(44)\,{\rm GeV},\label{eq:lat_scale}
\end{equation}
is 0.72\% smaller than the previous value of $a^{-1}=2.333(18)$ GeV, which was estimated with the physical inputs of $m_\pi$, $m_K$, and $m_\Omega$ on $96^4$ lattice in Ref.~\cite{k-config}. Since we find no reasonable plateau for the effective mass in the $\Omega$ baryon channel as shown in Sec.~\ref{subsec:fse_10b}, we should update the cutoff scale from the old estimate of $a^{-1}=2.333(18)$ GeV~\footnote{Although the old estimate is used in the previous work of Refs.~\cite{mf_nf2+1,nff_128,g-2_128}, we are not concerned with any practical problem because the old estimate is consistent with the new one within the error bar.}.

Having determined the physical point, we can now obtain the hadron masses at the physical point. We focus on the octet baryon masses, because it is impossible to extract the masses for the vector meson and the decuplet baryon sectors whose effective masses do not show any reasonable plateaus as found in Secs.~\ref{subsec:fse_v} and \ref{subsec:fse_10b}. We use the following formula to extrapolate the results to the physical point:
\ben
m_N&=&m_N\vert_{\rm sim}+\left.\frac{\partial m_N}{\partial m_{\rm ud}}\right\vert_{\rm sim}\left(m_{\rm ud}- m_{\rm ud}\vert_{\rm sim}\right)\nn\\
&&\hspace{15mm}+\left.\frac{\partial m_N}{\partial m_{\rm s}}\right\vert_{\rm sim}\left(m_{\rm s}- m_{\rm s}\vert_{\rm sim}\right),\\
m_\Lambda&=&m_\Lambda\vert_{\rm sim}+\left.\frac{\partial m_\Lambda}{\partial m_{\rm ud}}\right\vert_{\rm sim}\left(m_{\rm ud}- m_{\rm ud}\vert_{\rm sim}\right)\nn\\
&&\hspace{15mm}+\left.\frac{\partial m_\Lambda}{\partial m_{\rm s}}\right\vert_{\rm sim}\left(m_{\rm s}- m_{\rm s}\vert_{\rm sim}\right),\\
m_\Sigma&=&m_\Sigma\vert_{\rm sim}+\left.\frac{\partial m_\Sigma}{\partial m_{\rm ud}}\right\vert_{\rm sim}\left(m_{\rm ud}- m_{\rm ud}\vert_{\rm sim}\right)\nn\\
&&\hspace{15mm}+\left.\frac{\partial m_\Sigma}{\partial m_{\rm s}}\right\vert_{\rm sim}\left(m_{\rm s}- m_{\rm s}\vert_{\rm sim}\right),
\een
where the coefficients are determined by using the reweighted results on the $64^4$ lattice. The fit values of the coefficients are listed in Table~\ref{tab:fit_64}. Table~\ref{tab:fitmass_8b_pp} summarizes the results for the octet baryon masses at the physical point. They are consistent with those at the simulation point in Table~\ref{tab:fitmass_8b} within the error bars as expected from Fig.~\ref{fig:physicalpoint}.  
Note that our result for the nucleon mass is heavier than the experimental value by 1.25(45)\% beyond the error bar. There may be two possible reasons: the scaling violation effect and/or the lack of isospin breaking effect. We leave the investigation to future studies.

\begin{table}[t!]
\caption{Fit results for the AWI quark mass dependence of $m_\pi^2$, $m_K^2$, $m_\Xi$ and other octet baryon masses on $64^4$ lattice.}
\begin{center}
\begin{tabular}{ccc}
\hline
Mass  & $C_{\rm ud}$ & $C_{\rm s}$  \\\hline
$m_{\pi}^2$ & 2.445(21) & 0.0030(28)\\
$m_{K}^2$   & 1.362(89) & 1.206(14)\\
$m_{\Xi}$   & 3.6(1.8)  & 4.11(21)\\\hline
$m_N$       & 23.2(8.5) & 0.16(84)\\
$m_\Lambda$ & 14.5(5.5) & 2.43(41)\\
$m_\Sigma$  &  7.7(2.6) & 3.15(29)\\\hline
\end{tabular}
\end{center}
\label{tab:fit_64}
\end{table}%

\begin{table}[t!]
\caption{Bare AWI quark masses in the lattice unit and lattice cutoff scale at the physical point determined on $128^4$ lattice.}
\begin{center}
\begin{tabular}{cccc}
\hline
  & $m_{\rm ud}$ & $m_{\rm s}$ & $a^{-1}$ [GeV]  \\\hline
simulation point  & 0.0013663(143) & 0.037983(006) & $\cdots$\\
physical point    & 0.0013592(062) & 0.038048(147) & 2.3162(44)\\\hline
\end{tabular}
\end{center}
\label{tab:physicalpoint}
\end{table}%

\begin{table}[t!]
\caption{Octet baryon masses at the physical point on $128^4$ lattice.}
\begin{center}
\begin{tabular}{ccccccc}
\hline
Mass & lattice unit & Physical unit [GeV] & Experiment [GeV] \\\hline
$m_{N}$       & 0.4108(14) & 0.9514(30) & 0.9396\\
$m_{\Lambda}$ & 0.4828(16) & 1.1184(21) & 1.1157\\
$m_{\Sigma}$  & 0.5168(11) & 1.1971(15) & 1.1926 \\ \hline
\end{tabular}
\end{center}
\label{tab:fitmass_8b_pp}
\end{table}%

\section{Conclusions and outlook}
\label{sec:conclusion}

We have investigated the finite size effect on the vector meson and the baryon sectors using (5.5 fm$)^4$ and (10.9 fm$)^4$ lattices in 2+1 flavor QCD at the physical point. The effective masses for the vector meson sector do not show any reasonable plateau region so that we cannot extract the masses by the single exponential fit of their propagators. This is also true for the decuplet baryon sector. Both sectors show ``unstablelike'' behaviors on the lattice because of two common reasons. One is that there exist two-body states whose energy is close to that of the ground state. The other is that the $^2O$ representations for these sectors allow the mixing between the ground states and the small $\Delta j$ excited states. It is no wonder that the $\Omega$ baryon mass in the decuplet sector shows no plateau behavior, which is confirmed with high level of precision by varying the solver algorithm and the smearing parameter. 
The proper method to obtain the $\Omega$ baryon mass must be an analysis by the use of appropriate multiple states. 
On the other hand, the octet baryon sector has ``stable'' ground states on the lattice. Their masses do not show any finite size effect beyond 0.5\% level of statistical errors.
The choice of $m_\pi$, $m_K$, and $m_\Xi$ as the physical inputs yields the cutoff scale of $a^{-1}=2.3162(44)$ GeV together with the bare quark masses at the physical point. 
We find that the nucleon mass is heavier than the experimental value by 1.25(45)\% beyond the error bar. This could be due to the cutoff effect and/or the lack of the isospin breaking effect. We plan to investigate it using the PACS10 configurations at finer lattice spacings.

\begin{acknowledgments}
Numerical calculations are carried out on Oakforest-PACS through the HPCI System Research project (Project No.~hp170093, No.~hp180051, No.~hp190081, No.~hp190068) and the Interdisciplinary Computational Science Program in Center for Computational Sciences, University of Tsukuba. This work is supported in part by Grants-in-Aid for Scientific Research from the Ministry of Education, Culture, Sports, Science and Technology (MEXT) (Grants No.~16K13798, No.~16H06002, No.~18K03638, No.~19H01892). 
\end{acknowledgments}

\end{document}